\documentclass{revtex4}
\usepackage{amssymb,amsmath,upgreek,bm} 
\usepackage{xcolor}
\usepackage{hyperref}
\hypersetup{linkbordercolor=pink}
\usepackage{graphicx}
\DeclareGraphicsExtensions{.ps,.eps}
\usepackage{epstopdf}
\newcommand{\beq}{\begin{equation}}
\newcommand{\eeq}{\end{equation}}
\newcommand{\beqa}{\begin{eqnarray}}
\newcommand{\eeqa}{\end{eqnarray}}
\newcommand{\beqan}{\begin{eqnarray*}}
\newcommand{\eeqan}{\end{eqnarray*}}
\newcommand{\bra}[1]{\langle{#1}|}

\newcommand{\ket}[1]{|{#1}\rangle}

\newcommand{\ip}[1]{\langle{#1}\rangle}

\newcommand{\eq}[1]{Eq.~\eqref{#1}}
\newcommand{\sfc}[2]{\mbox{$\frac{#1}{#2}$}}
\newcommand{\srel}[2]{{\def\arraystretch{0.25}\begin{array}{cc}\\#1 \\ #2\end{array}}}

\begin{document}
	
\title{Quantum asymmetry between time and space}

\author{Joan A. Vaccaro}

\address{Centre for Quantum Dynamics, Griffith University, Nathan 4111 Australia}
	
%
%

\begin{abstract} 

An asymmetry exists between time and space in the sense that physical systems inevitably evolve over time whereas there is no corresponding ubiquitous translation over space.  The asymmetry, which is presumed to be \textit{elemental}, is represented by equations of motion and conservation laws that operate differently over time and space.  If, however, the asymmetry was found to be due to deeper causes, this conventional view of time evolution would need reworking.  Here we show, using a sum-over-paths formalism, that a violation of time reversal (T) symmetry might be such a cause. If T symmetry is obeyed, the formalism treats time and space symmetrically such that states of matter are localised both in space and in time. In this case, equations of motion and conservation laws are undefined or inapplicable. However if T symmetry is violated, the same sum over paths formalism yields states that are localised in space and distributed without bound over time, creating an asymmetry between time and space. Moreover, the states satisfy an equation of motion (the Schr\"{o}dinger equation) and conservation laws apply. This suggests that the time-space asymmetry is \textit{not elemental} as currently presumed, and that T violation may have a deep connection with time evolution.

\end{abstract}

\maketitle

	
\section{Introduction}

There is nothing unphysical about matter being localised in a region of space; matter can simply exist at one location and not another.  But for it to be localised in a finite period of time is altogether different.  Indeed, as the matter would exist only for that period and no other, the situation would be a direct violation of mass conservation.
In conventional quantum mechanics, this undesirable situation is avoided axiomatically by requiring matter to be represented by a quantum state vector whose norm is fixed over time.  There is, however, no corresponding restriction of the state vector over space.



\noindent

The underlying time-space asymmetry here can be traced to the fact that the state, and the matter it represents, is presumed to undergo continuous translation over time (as time evolution) but there is no corresponding presumption about the state undergoing translations over space.  Even in relativistic quantum field theory, where both time and space are treated equally as the coordinates of a spacetime background, a similar asymmetry holds because time evolution and conservation laws are presumed to operate differently over time and space.

Nevertheless, time and space could have an equivalent footing at a fundamental level if any asymmetry between them were to arise \textit{phenomenologically} rather than being imposed axiomatically.  Such a prospect is well worth pursuing because it would help us to understand the relationship between time and space.  It would require finding an underlying mechanism that, due to phenomenological conditions, affects the spatial and temporal translational degrees of freedom in different ways to the extent matter can be localised in space but not in time. This suggests that we should examine the phenomenological character of the operations associated with the translational degrees of freedom.
The generators of translations in space and time are given by the momentum and Hamiltonian operators, respectively, and with them lies a difference that sets space and time apart in the quantum regime. 

In fact, the last fifty years \cite{parity,CP,T1,T2,T3,T4} has shown that Nature is not invariant to particular combinations of the discrete symmetry operations of charge conjugation (C), parity inversion (P) and time reversal (T). The violation of these discrete symmetries are observed in various particle decays independent of position in space, and so they occur over translations in time and not translations in space. In terms of the corresponding generators, this implies that the Hamiltonian violates the discrete symmetries whereas the momentum operator does not.

The discrete symmetry violations are accounted for in the Standard Model of particle physics by the Cabibbo-Kobayashi-Maskawa (CKM) matrix \cite{Cabibbo,Kobayashi-Maskawa}.  
Studies of the violations have been made in relation to baryogenesis in the early universe \cite{Sakharov}, the arrow of time and irreversibility \cite{Aharony1,Aharony2,Gell-Mann,Berger}, the time operator \cite{Courbage2},  quantum entanglement and Bell inequalities \cite{Datta1,Finkelstein,Hall,Lindblad,Squires,Datta2,Corbett,Clifton,Bramon,Ancochea,Bertlmann1,Bertlmann2,Andrianov,Barnett,Samal,Bertlmann4,Genovese}, decay and decoherence \cite{Courbage1,Fassarella},  complementarity and quantum information \cite{Bertlmann3,Bertlmann5,Hiesmayr}, quantum walks \cite{Lu}, and the potential for T violation to have large scale physical effects \cite{FPhys,FPhys2}. In particular, in Ref.~\cite{FPhys} I modelled the state of the universe as a superposition of paths that zigzag through time, and showed that T violation can, in principle, affect the time evolution in a global way. Then in Ref.~\cite{FPhys2} I showed that the effect on time evolution is greater when the paths are constructed in the limit of infinitely-small steps. 

Here we explore the potential impact the violations of the discrete symmetries may have for giving quantum states different representations in space and time.  The aim is not to study specific instances of the violations as observed experimentally, but rather to look for possible consequences of the violations in general terms.  For this, the definitions of the P and T operations given by Wigner in relation to non-relativistic quantum mechanics \cite{Wigner} are sufficient and so we shall undertake the analysis using the same theory as a basis. Many previous studies have used the same framework \cite{Aharony1,Aharony2,Gell-Mann,Berger,Courbage2,Datta1,Finkelstein,Hall,Lindblad,Squires,Datta2,Corbett,Clifton,Bramon,Ancochea,Bertlmann1,Bertlmann2,Andrianov,Barnett,Samal,Bertlmann4,Genovese,Courbage1,Fassarella,Bertlmann3,Bertlmann5,Hiesmayr,Lu}.
A relativistic analysis will be left for a future study.

We will need, however, to depart from conventional quantum mechanics in three important ways. The justification for these departures lies in the eventually recovery of the conventional formalism under appropriate conditions. The first departure is that we will not impose any equation of motion, such as the Schr\"{o}dinger equation, on states because to do so would directly build in the asymmetry between time and space mentioned above. Instead, we anticipate that an effective equation of motion will arise phenomenologically in some way.  Second, we will consider states that describe the location of a material object either in space or in time.  While the location in space can simply be given by a wave function, say $\psi(\vec{x})$, in the position representation, the location in time is quite unconventional as it would need to be given by a wave function, say $\phi(t)$, in what might be called the ``time representation''. Here, $|\phi(t)|^2$ gives the probability density for the object being at time $t$ just as $|\psi(\vec{x})|^2$ gives the probability density for the object being at position $\vec{x}$.  Note that, in general, $\phi(t)$ violates mass conservation as it describes the object as having potential existence at the points in time where $|\phi(t)|^2\ne0$ and definitely not existing at the points where $|\phi(t)|^2=0$. This lack of mass conservation is the price we must pay to keep the formalism symmetric with respect to time and space.  Nevertheless we anticipate that mass conservation will arise phenomenologically rather than being imposed on the formalism. The third departure from conventional quantum mechanics is that we need to include the P and T symmetry operations in the formalism explicitly, even in the situation where the corresponding symmetries are obeyed and the actions of P and T are redundant. This will ensure that we have a consistent formalism that operates both when the symmetries are obeyed and also when they are violated.

The particular way in which we include the P and T discrete symmetry operators is motivated by the fact that they reverse the directions of spatial and temporal translations, respectively. A one-dimensional path in space that consists of a sequence of translations that alternate in direction can, therefore, be expressed explicitly in terms of parity inversions P. The same can be said for the T operation in relation to a sequence of time translations that alternate in direction. The roles that P and T play will be greater in mathematical constructions that involve a greater number of direction reversals.  Such constructions have the potential to display the effects of any discrete symmetry violation to a greater extent. Feynman's path integral formalism immediately comes to mind as one that involves a superposition of all paths that zigzag through configuration space between two states. However, this formalism is inextricably associated with dynamics and thus is tied to the space-time asymmetry mentioned above.  We need to develop a different approach if we are to keep time and space on an equal footing at a fundamental level.

Our approach is as follows.  We will first consider the effects of P violation on spatial localisation.
We will begin with a quantum state that represents some material object as being localised in space, and for this we will need the variance in the object's position to be finite.
We will place no other constraint on the position and so we will need a quantum state that yields the least information about position (and thus maximum entropy) for a fixed variance; the optimum pure state fitting this requirement has a Gaussian wave function in the position representation \cite{Bialynicki-Birula}. Note that a classical particle undergoing a one-dimensional Wiener process has a position probability density that is Gaussian; it too has trajectories that consist of infinitely many reversals in direction of the kind we have been considering. With this in mind, we will decompose the Gaussian quantum state into a superposition of  infinitely-many paths through space where each path has infinitely-many reversals in direction. The reversals in the direction of each path will be expressed explicitly in terms of the P symmetry operation and the translations in terms of the momentum operator. We will find that the violation of P symmetry has no effect on the construction. Then, we will apply the same sum-over-paths construction to a quantum state that represents the object as being localised in time, but with the path reversals expressed explicitly in terms of the T symmetry operator and the translations in terms of the Hamiltonian.  Using the same construction will ensure that \textit{the formalism is symmetric with respect to the representation of states of matter in both space and time} when the discrete symmetries hold.  The situation will be found to change dramatically when T symmetry is violated.  Only then will the formalism exhibit a time-space asymmetry that is consistent with conventional quantum mechanics.  The important point to be made here is that the asymmetry will not be imposed on the formalism at a fundamental level, but rather it will arise \textit{phenomenologically} due to the T violation.

Given the fundamental character of the issues involved, one should not be surprised to find that to make any progress we need to pay due attention to quite subtle mathematical details.  In particular, while the concept of the limit of an infinite sequence has rigorous meaning in a mathematical context, there is no \textit{a priori} reason to suppose that it automatically carries a corresponding value in a theory that is designed to underpin experimental physics.  After all, the accuracy of observations made in experimental physics are always restricted by finite resources.  For example, consider a theory in which the limit point $a$ of the convergent sequence $a_1$, $a_2$, $a_3$, $\ldots$ (i.e. where $a_n\to a$ as $n\to\infty$) represents an experimental parameter, and let $\epsilon$ represent the experimental accuracy of measuring $a$ for a given level of resources.  The convergence of the sequence implies that there exists a natural number $N_\epsilon$ that depends on $\epsilon$ for which $|a-a_n|<\epsilon$ for all $n>N_\epsilon$, and so it is not possible to physically distinguish (using the given resources)  the limit point $a$ from any of the terms $a_n$ for $n>N_\epsilon$.   Under such circumstances, the set  $\{a_n:n>N_\epsilon\}$ would be more  representative of the physical situation than just the limit point $a$. Set representations of this kind will be important for expressing quantum states in a manner which better represents their physical implications.

The structure of the remainder of the paper is as follows.  We develop a sum-over-paths construction of a quantum state that is localised in space and examine the effects of the violation of P symmetry in section \ref{sec:construction}. We apply the same construction to quantum states that are localised in time and examine the effects of the violation of T symmetry in section \ref{sec:applied to time}. Following that, in section \ref{sec:emergence}, we show how the conventional Schr\"{o}dinger equation and conservation of mass emerge as a result of coarse graining over time, and explore how the new formalism might be tested experimentally.  We end with a discussion in section \ref{sec:conclusion}.

\section{\label{sec:construction}Mathematical construction of quantum states in space}

\subsection{Developing the construction}

\begin{figure}  
	\begin{center}
		\includegraphics[width=8.0cm]{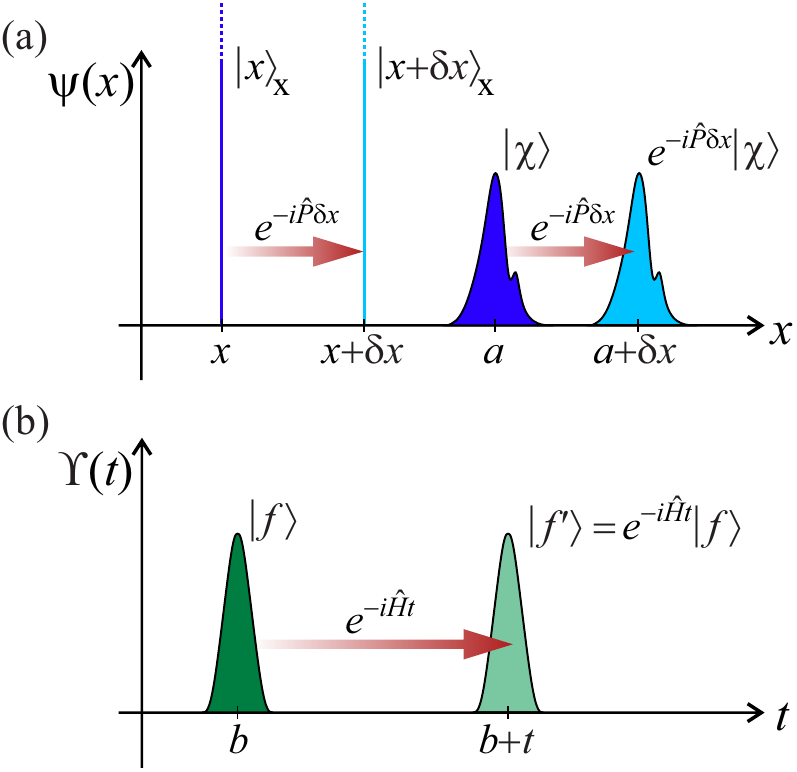}
	\end{center}
	\vspace{-5mm}
	\caption{Sketches illustrating the translation of wave functions along (a) the $x$ axis and (b) the time axis.  In (a) the wave functions represent the position eigenket $\ket{x}_{\rm x}$ and an arbitrary state $\ket{\chi}$ and the translation is by a distance $\delta x$.  In (b) the wave function represents the state $\ket{f}$ and the translation is by an interval $t$.
		\label{fig:translation}}
\end{figure}

We first need to develop the mathematical construction of quantum states that are localised in space and consist of a superposition of infinitely-many paths each of which has possibly infinitely-many reversals in direction.  For this consider a simple 1-dimensional model universe composed of a single ``galaxy'' as our material object. The galaxy is representative of any spatially localised physical system with mass and could in fact be a star, a planet or just a single particle; its details are not important for this study. The location of the galaxy is described by a set of observables that represent all its spatial degrees of freedom.  Imagine that at a particular time, each observable is in some localised state that is uncorrelated with respect to every other observable in the set.  This will almost certainly result in the galaxy being far from its minimum energy state, however neither the energetics nor the dynamics are important in this section. Also, because the same analysis applies to each observable, we will only treat one representative observable explicitly. Let that observable be the centre of mass coordinate, which we assume to have a finite variance.   As mentioned above, the best choice for a pure state under these circumstances is one described by a Gaussian wave function \cite{Bialynicki-Birula}, which we write as follows:
\begin{equation} \label{eq:spatial gaussian state}
     \ket{\psi} \propto \int dx \exp(-\frac{x^2}{2\sigma^2_{\rm x}})\ket{x}_{\rm x}
\end{equation}
where  $x$ and $\ket{x}_{\rm x}$ are the eigenvalue and corresponding eigenstate of $\hat X$, the operator representing the $x$ component of the centre of mass position, and $\sigma_{\rm x}$ is a width parameter.
This state can be written explicitly in terms of spatial translations as
\begin{equation}   \label{eq:gaussian spatial translation}
    \ket{\psi}  \propto \int dx \exp(-\frac{x^2}{2\sigma^2_{\rm x}})\exp(-i\hat Px)\ket{0}_{\rm x}
\end{equation}
where operator representing the total momentum of the galaxy, $\hat P$,  generates spatial translations according to
\[
 \exp(-i\hat P\delta x)\ket{x}_{\rm x}=\ket{x+\delta x}_{\rm x}
\]
as illustrated in Fig.~\ref{fig:translation}(a).   Here, and throughout this paper, we use units in which $\hbar=1$.
Inserting the resolution of the identity $\hat{\boldsymbol 1}=\int dp \ket{p}_{\rm p}{}_{\rm p}\bra{p}$ into \eq{eq:gaussian spatial translation} gives
\[
\ket{\psi}  \propto \iint dx\,dp \exp(-\frac{x^2}{2\sigma^2_{\rm x}})\exp(-ipx)\ket{p}_{\rm p}{}_{\rm p}\ip{p|0}_{\rm x}\ ,
\]
where $\{\ket{p}_{\rm p}: \hat P\ket{p}_{\rm p}=p\ket{p}_{\rm p}\}$ is the momentum basis.  Carrying out the Fourier transform with respect to $x$, yields
$\ket{\psi}  \propto \exp(-\sfc{1}{2}\hat P^2\sigma^2_{\rm x})\ket{0}_{\rm x}$ and making use of the result
\begin{equation}
\exp(-A^2/2)=\lim_{N\to\infty}\cos^N(A/\sqrt{N})
\end{equation}
then leads to
\begin{equation} \label{eq:path in space}
   \ket{\psi}  \propto
     \lim_{N\to\infty}\frac{1}{2^N}\!\left[\exp(i\frac{\hat P\sigma_{\rm x}}{\sqrt N})+\exp(-i\frac{\hat P\sigma_{\rm x}}{\sqrt N})\right]^N\kern-1mm\ket{0}_{\rm x}\ .
\end{equation}

\begin{figure}  
	\begin{center}
		\includegraphics[width=8.0cm]{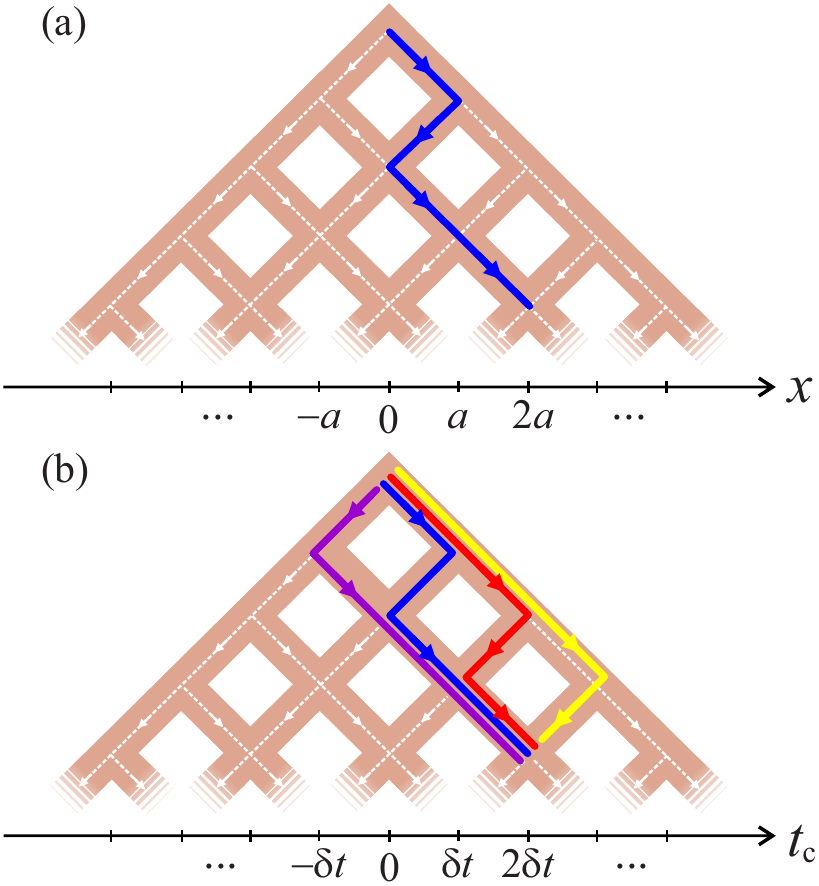}
	\end{center}
	\vspace{-8mm}
	\caption{Binary tree diagrams representing virtual paths in (a) space and (b) time.  Each edge (white dashed line) in the tree represents a virtual displacement along the black horizontal axis.  The thick blue edges in (a) represents a virtual path that passes through the sequence of points $0$, $a$, $0$, $a$, $2a$ on the $x$ axis.  In (b) four different virtual paths from $0$ to $2\delta t$ on the $t_{\rm c}$ axis are represented in the tree by thick edges coloured yellow, red, blue and purple.
		\label{fig:random path}}
\end{figure}

Expanding the $N$-fold product in \eq{eq:path in space} gives a series of terms each of which comprise $N$ translations (or ``steps'') of $\pm\sigma_{\rm x}/\sqrt{N}$ along the $x$ axis.
For example, a term of the form
\begin{equation} \label{eq:one path}
 \cdots\exp(-i\hat P a)\exp(-i\hat P a)\exp(i\hat P a)\exp(-i\hat P a)\ket{0}_{\rm x}\ ,
\end{equation}
where $a=\sigma_{\rm x}/\sqrt{N}$, describes a path on the $x$ axis from the origin $0$ through the sequence of points $a$, $0$, $a$, $2a$ and so on, as illustrated in Fig.~\ref{fig:random path}(a). Equation \eq{eq:path in space} can be viewed, therefore, as a superposition of random paths away from the origin $\ket{0}_{\rm x}$ in the limit of infinitely small steps, and shares similarities with both quantum walks \cite{quantum-walk} and Feynman's sum over paths \cite{Feynman}.
Note that here, however, the random path is traversed without reference to time, and so it should be considered to be traversed in a zero time interval. Each random path is, therefore, a generalisation of the virtual displacements in D'Alembert's principle in classical mechanics \cite{Goldstein}.  For this reason each individual path shall be called a {\it random virtual path} and the superposition of a set of random virtual paths like that in \eq{eq:path in space} shall be called a {\it quantum virtual path}.

Although they share similarities, a quantum virtual path is quite distinct from Feynman's sum over paths \cite{Feynman}.  For example,  Feynman's method is used to calculate the probability amplitude for a system to evolve from one state to another.  The paths represent potential classical trajectories between the same starting and ending points and the sum gives the total probability amplitude for evolving between the points.  In contrast, a quantum virtual path represents a single state. The accumulated displacement over one random virtual path, like that in \eq{eq:one path}, gives a potential classical position of the system, and the whole quantum virtual path represents the state \eq{eq:path in space} in terms of a distribution of potential classical positions. Moreover, calculating the innerproduct of two states where one (or both) is represented by a quantum virtual path would result in a Feynman-like sum over paths calculation.  So in this sense, a quantum virtual path is a \textit{precursor} of Feynman's sum over paths.

The right-hand side of \eq{eq:path in space} is not the only way to decompose the state in \eq{eq:spatial gaussian state}.  But what makes \eq{eq:path in space} special is that it consists of a superposition of an infinite number of continuous paths with the property that if one path is picked at random, it will effectively consist of a sequence of infinitesimal segments where each segment has an equal likelihood of representing a step in the positive or negative $x$ directions.  The set of paths is unbiased with respect to direction in this sense. Another feature that sets the decomposition in \eq{eq:path in space} apart is that it comprises all possible paths. The justification of why it should is that in decomposing the state in \eq{eq:spatial gaussian state} in terms of a superposition of paths, we have no reason for leaving out particular paths or, alternatively, for including only particular paths; in the absence of such reasons all possible paths should be included.  

As $N\to\infty$ the step length $\sigma_{\rm x}/\sqrt{N}$  in \eq{eq:path in space} will eventually breach the fundamental lower bound, say  $\delta x_{\rm min}$, that is expected for physically distinguishable positions.
For example, there are reasons \cite{Planck} to believe that points in space are indistinguishable at the scale of the Planck length $\ell_{\rm P}\approx 1.6\times 10^{-35}$~m.
Let $N^{\rm (space)}_{\rm min}$ be the value of $N$ where the step length $\sigma_{\rm x}/\sqrt N$ becomes equal to $\delta x_{\rm min}$, i.e. $N^{\rm (space)}_{\rm min}=\sigma^2_{\rm x}/\delta x_{\rm min}^2$.  This implies that the limit on the right side of \eq{eq:path in space} can be replaced by a term corresponding to any value of $N$ larger than $N^{\rm (space)}_{\rm min}$ without any physically meaningful consequences.  There are an infinite number of such terms, each of which has an \textit{equal status} in representing the state of the universe.
They form the set
\begin{equation}  \label{eq:set Psi}
   \boldsymbol{\Uppsi}=\{\ket\psi_N: N\ge N^{\rm (space)}_{\rm min}\}
\end{equation}
where
\begin{equation} \label{eq:psi_N}
   \ket\psi_N=
     \frac{1}{2^N}\!\left[\hat{\mathbf{P}}^{-1}\!\exp(-i\frac{\hat P\sigma_{\rm x}}{\sqrt N})\hat{\mathbf{P}}+\exp(-i\frac{\hat P\sigma_{\rm x}}{\sqrt N})\right]^N\kern-1mm\ket{0}_{\rm x}\ .
\end{equation}
In \eq{eq:psi_N} we have written the translations explicitly in terms of the parity inversion operator $\hat{\mathbf{P}}$.  It has the property that
\begin{equation}   \label{eq:parity op for step}
   \exp(i\hat P x')=\hat{\mathbf{P}}^{-1}\!\exp(-i\hat P x')\hat{\mathbf{P}}
\end{equation}
which expresses the fact that a translation along the $x$ axis by $-x'$ (left side of \eq{eq:parity op for step}) can be produced by first performing a parity inversion, translating by $x'$ and then reversing the parity inversion (right side).
Every element in the set $\boldsymbol{\Uppsi}$ can serve equally well as a representation of the state in \eq{eq:spatial gaussian state} as far as the physically-distinguishable spatial limit allows;
they all have equal status in this respect.

The mathematical construction represented by \eq{eq:set Psi} and \eq{eq:psi_N} is in the form of the explicit translations and discrete symmetry operations that we need for comparing the difference between quantum states in space and time.
Although being equivalent to \eq{eq:spatial gaussian state}, we shall henceforth regard \eq{eq:set Psi} and \eq{eq:psi_N} as being a \textit{more fundamental} description of the state of the galaxy due to this explicit form.
Note that the interpretation of \eq{eq:psi_N} in terms of quantum virtual paths does not hinge on the state $\ket{0}_{\rm x}$ being the eigenstate of position with zero eigenvalue.  In fact any state $\ket{\chi}$ with a variance in position very much smaller than $\sigma^2_{\rm x}/2$ (and, correspondingly, a variance in total momentum very much larger than $1/2\sigma^2_{\rm x}$) could be used in its place, in which case the steps in a path represent translations of $\ket{\chi}$ along the $x$ axis, as illustrated in Fig.~\ref{fig:translation}(a), rather than steps along the $x$ axis itself.  While this situation allows some ambiguity in the formalism, it does not have any effect on results provided that corresponding adjustments to $\ket{\psi}$ and $\sigma_{\rm x}$ are duly taken.

\subsection{Parity inversion invariance and its violation}

Given that the Hamiltonian does not appear explicitly in the construction of spatial states, we should not expect to find any effects of discrete symmetry violation here. In particular, \textit{regardless of whether the galaxy obeys parity inversion symmetry or not}
\begin{equation}
     \hat{\mathbf{P}}^{-1}\hat P\,\hat{\mathbf{P}}=-\hat P
\end{equation}
always holds, and so \eq{eq:psi_N} can be rewritten as a binomially weighted superposition of spatially translated states, i.e.
\begin{equation} \label{eq:spatial binomial state}
    \ket\psi_N
     = \sum_{n=0}^N B_n \exp\left[-i(2n-N)\frac{\hat P\sigma_{\rm x}}{\sqrt N}\right]\ket{0}_{\rm x}
\end{equation}
where
\begin{equation} \label{eq:binomial coefficients}
    B_n =
    \frac{1}{2^N}\left({}^N_n\right)\ .
\end{equation}
In the large $N$ limit, $\ket\psi_N$ tends to the Gaussian state $\ket{\psi}$ in \eq{eq:spatial gaussian state}, i.e.
\begin{equation} 
     \lim\limits_{N\to\infty}\ket\psi_N  \propto \int dx\  g(x,\sigma_{\rm x}) \ket{x}_{\rm x}
\end{equation}
where
\begin{equation} \label{eq:gaussian function}
     g(x,\sigma_{\rm x}) = \exp(-\frac{x^2}{2\sigma^2_{\rm x}})\ .
\end{equation}

\begin{figure}  
 	\begin{center}
  		\includegraphics[width=9.0cm]{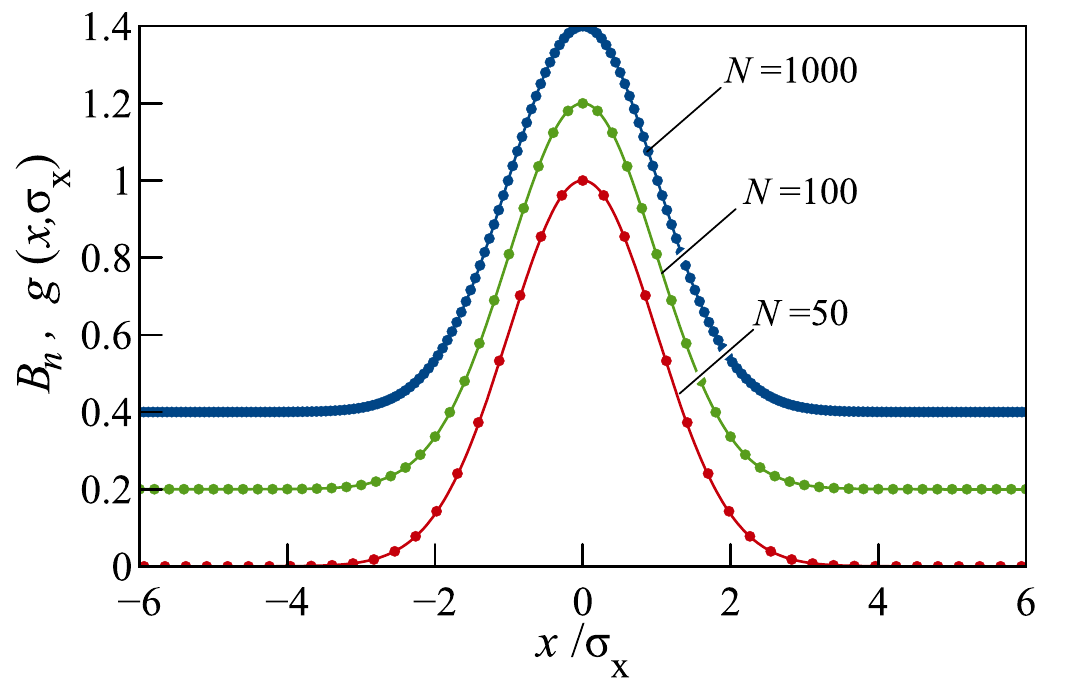}
 	\end{center}
 	\vspace{-8mm}
 	\caption{The position representation of the state $\ket{\psi}_N$ as a function of the scaled position $x/\sigma_{\rm x}$. The dots represent the exact values $B_n$ given by \eq{eq:binomial coefficients} and the solid curves represent the Gaussian approximation  $g(x,\sigma_{\rm x})$ given by \eq{eq:gaussian function}. The abscissae for the discrete coefficients $B_n$ are given by $x/\sigma_{\rm x}=(2n-N)/\sqrt{N}$ in accord with \eq{eq:spatial binomial state}. For clarity, $B_n$ and $g(x,\sigma_{\rm x})$ have been scaled to give a maximum of unity, and the green ($N=100$) and blue ($N=1000$) dots and curves have been displaced vertically by 0.2 and 0.4, respectively.  
	 	\label{fig:TInv}}
\end{figure}    

Figure \ref{fig:TInv} compares the coefficients $B_n$ (shown as dots) with their large-$N$ limit $g(x,\sigma_{\rm x})$ (continuous curves) for a number of different $N$ values.  The values of $N$ have been chosen purposely to exaggerate the discreteness of the state $\ket{\psi}_N$ in comparison to the limiting state  $\ket{\psi}$ from  \eq{eq:spatial gaussian state}. In truth, for every $\ket{\psi}_N\in \boldsymbol{\Uppsi}$ in \eq{eq:set Psi}, the values of $N$ are sufficiently large (viz. $N\ge N^{\rm (space)}_{\rm min}$) that the dots representing $B_n$ for consecutive $n$ values are physically indistinguishable, and the locus of points representing $B_n$ is essentially equivalent to the curve $g(x,\sigma_{\rm x})$ up to a proportionality constant.  As a consequence, every $\ket{\psi}_N\in \boldsymbol{\Uppsi}$ is physically indistinguishable from the state $\ket{\psi}$ in \eq{eq:spatial gaussian state}.

\section{\label{sec:applied to time}Applying the construction to quantum states in time}

\subsection{Adapting the construction}

We now use our construction to explore the temporal analogy of  \eq{eq:spatial gaussian state} in which the galaxy is represented in time rather than space.
We begin by recalling that the Hamiltonian $\hat H$ generates translations through time according to
\[
    \exp(-i\hat H t)\ket{f}=\ket{f'}
\]
where $\ket{f}$ and $\ket{f'}$ represent states at times differing by $t$, as illustrated in Fig.~\ref{fig:translation}(b).
Next, we construct a set of states analogous to \eq{eq:set Psi} but with each state representing a superposition of random virtual paths through {\it time} as
\begin{equation}    \label{eq:set Upsilon}
     \boldsymbol{\Upupsilon}_\lambda =\{ \ket{\Upsilon_\lambda}_N: N\ge N^{\rm (time)}_{\rm min}\}
\end{equation}
where
\begin{equation} \label{eq:Upsilon_N}
    \ket{\Upsilon_\lambda}_N \propto \frac{1}{2^N}\!\left[\hat{\mathbf{T}}^{-1}\!\exp(-i\frac{\hat H\sigma_{\rm t}}{\sqrt N})\hat{\mathbf{T}}+\exp(-i\frac{\hat H\sigma_{\rm t}}{\sqrt N})\right]^N\kern-1mm\ket{\phi}\ .
\end{equation}
Here $\lambda$ distinguishes different physical situations that will be specified later, $N^{\rm (time)}_{\rm min}=\sigma_{\rm t}^2/\delta t_{\rm min}^2$ is the value of $N$ for which the step size $\sigma_{\rm t}/\sqrt{N}$ reaches some fundamental resolution limit in time $\delta t_{\rm min}$ (e.g. taking the resolution limit as the Planck time would mean that $\delta t_{\rm min}=5.4\times 10^{-44}$~s), and $\hat{\mathbf{T}}$ is Wigner's  time reversal operator \cite{Wigner}.
The state $\ket{\phi}$ plays the role of $\ket{0}_{\rm x}$  in \eq{eq:psi_N} and is assumed to be sharply defined in time and, correspondingly, to have a broad distribution in energy \cite{Pegg}.  More specifically, $\ket{\phi}$ must have a variance in energy that is very much larger than $1/2\sigma_{\rm t}^2$ in analogy with the requirement for any state $\ket{\chi}$ to be used in place of $\ket{0}_{\rm x}$.  Other details of $\ket{\phi}$ are not crucial for our main results.   

It is perhaps worth elaborating a little on what is meant by  $\ket{\phi}$  being sharply defined in time given that there are well-known difficulties associated with defining an operator to represent time \cite{Moyer,Pegg}.  Fortunately, the absence of a universally accepted time operator does not prevent uncertainties in time from being physically meaningful.  Rather, we can use the fact that the Hamiltonian is the generator of translations in time to probe the time uncertainty of a state. For example,  $\ket{\varphi_\tau}=\exp(-i\hat H\tau)\ket{\varphi_0}$ represents the state $\ket{\varphi_0}$ translated in time by $\tau$.  If the overlap $\ip{\varphi_\tau|\varphi_0}$ is negligible for all values of $\tau$ except for $|\tau|\approx 0$ then $\ket{\varphi_0}$ can be regarded as sharply defined in time, at least for the purposes needed here.  A more rigorous definition of such states is given by Moyer's timeline states \cite{Moyer}.  The ambiguity mentioned at the end of the previous section also occurs here regarding the choice of the state $\ket{\phi}$, and can be treated in a similar way.  We will return to this point in section \ref{sec:emergence}\ref{sec:conventional qm}.

The violation of T symmetry is expressed by $\hat{\mathbf{T}}^{-1}\!\hat H\hat{\mathbf{T}}\ne\hat H$ which implies that there are two versions of the Hamiltonian \cite{FPhys,FPhys2}.  We label the two versions as $\hat H_{\rm F}=\hat H$ and $\hat H_{\rm B}=\hat{\mathbf{T}}^{-1}\!\hat H\hat{\mathbf{T}}$. 
In this construction, one direction of time is not physically distinguishable from the other and the subscripted labels ${\rm F}$ and ${\rm B}$ simply refer to opposite directions; nevertheless, it may be convenient to think of the labels as referring to the  customary ``forwards'' and ``backwards'' directions of time. 
Using these definitions together with the fact \cite{Wigner} that $\hat{\mathbf{T}}^{-1}i\hat{\mathbf{T}}=-i$ then gives
\begin{equation}   \label{eq:Upsilon_N in terms of H_F H_B}
     \ket{\Upsilon_\lambda}_N \propto \frac{1}{2^N}\!\left[\exp(i\hat H_{\rm B}\delta t)+\exp(-i\hat H_{\rm F}\delta t)\right]^N\kern-1mm\ket{\phi}
\end{equation}
where, for convenience, we have set 
\begin{equation}  \label{eq:time step}
    \delta t=\frac{\sigma_{\rm t}}{\sqrt{N}}
\end{equation}
as the step in time. 

\eq{eq:Upsilon_N in terms of H_F H_B} shows that $\hat H_{\rm F}$ and $\hat H_{\rm B}$ are responsible for translations in opposite directions of time. This is an important point that warrants particular emphasis: a translation in time in the opposite direction to that given by $\exp(-i\hat H_{\rm F} t)$ is not produced by its {\it inverse} $\exp(i\hat H_{\rm F} t)$ but rather by its {\it time reverse}:
\[
  \exp(i\hat H_{\rm B} t)=\hat{\mathbf{T}}^{-1}\! \exp(-i\hat H_{\rm F} t)\hat{\mathbf{T}}\ .
\]
Evidently we need to associate the operators $\exp(-i\hat H_{\rm F} t)$ and $\exp(i\hat H_{\rm B} t)$ with  {\it physical evolution} in different directions of time according to \eq{eq:Upsilon_N in terms of H_F H_B}.  This leaves their respective inverses $\exp(i\hat H_{\rm F} t)$ and $\exp(-i\hat H_{\rm B} t)$ to be associated with the {\it mathematical operations of rewinding} that physical evolution.  In short, physical time evolution is described by the former pair of operators, and not the latter.

In fact, these associated meanings follow from conventional quantum mechanics. For example, let $\ket{f(t)}$ represent the state of an arbitrary closed system at time $t$.  Unitary evolution implies that
\begin{equation}
   \ket{f(t)}=\exp(-i\hat h t)\ket{f(0)}
\end{equation}
where $\ket{f(0)}$ is the state at $t=0$ and $\hat h$ is the corresponding Hamiltonian. Recall that Wigner's time reversal operator $\hat{\mathbf{T}}$ reverses the direction of all momenta and spin \cite{Wigner}.   Let the time-reversed states at times $0$ and $t$ be $\ket{b(0)}=\hat{\mathbf{T}}^{-1}\ket{f(0)}$ and $\ket{b(-t)}=\hat{\mathbf{T}}^{-1}\ket{f(t)}$, respectively.
Using $\hat{\mathbf{T}}\hat{\mathbf{T}}^{-1}=\hat 1$ and rearranging shows that $\ket{b(-t)}=\exp(i\hat{\mathbf{T}}^{-1}\hat h\hat{\mathbf{T}} t)\hat{\mathbf{T}}^{-1}\ket{f(0)}$, i.e.
\begin{equation}
    \ket{b(-t)}=\exp(i\hat{\mathbf{T}}^{-1}\hat h\hat{\mathbf{T}} t)\ket{b(0)}\ ,
\end{equation}
and so the time-reversed state $\ket{b(-t)}$ represents the evolution from the time-reversed state $\ket{b(0)}$ according to the Hamiltonian $\hat{\mathbf{T}}^{-1}\hat h\hat{\mathbf{T}}$ for the time $-t$.
That is, evolving from the state $\ket{f(0)}$ for the time $t$ with the Hamiltonian $\hat h$ is equivalent to evolving from the time-reversed state $\ket{b(0)}$ for the time $-t$ with the Hamiltonian $\hat{\mathbf{T}}^{-1}\hat h\hat{\mathbf{T}}$. Clearly $\hat h$ generates translations in one direction of time and $\hat{\mathbf{T}}^{-1}\hat h\hat{\mathbf{T}}$ generates translation is the opposite direction, which is consistent with \eq{eq:Upsilon_N in terms of H_F H_B}.

If our model universe satisfied T symmetry, $\hat H_{\rm F}$ and $\hat H_{\rm B}$ would be commuting operators and the terms in \eq{eq:Upsilon_N in terms of H_F H_B} would be able to be manipulated algebraically in exactly the same way as those in \eq{eq:psi_N} led to \eq{eq:spatial binomial state}.  Thus, for the temporal quantum virtual path to be qualitatively distinct from the spatial one, the model universe must violate T symmetry to the extent of giving a non zero commutator $[\hat H_{\rm F},\hat H_{\rm B}]$.  We could model such a commutator using details of the T violation that has been observed in the decay of mesons \cite{T1,T2,T3,T4} or that has been speculated for a Higgs field \cite{Ivanov,Lee}. However, the potential repercussions of T violation will be manifest most clearly for the simplest departure from time reversal invariance. Accordingly we shall imagine that our model universe contains an unspecified T-violating mechanism that is consistent with the commutator 
\begin{equation}   \label{eq:commutator}
   [\hat H_{\rm B},\hat H_{\rm F}]=i\lambda
\end{equation}
for real valued $\lambda$. This is the origin of the parameter $\lambda$ that appears in \eq{eq:set Upsilon} and \eq{eq:Upsilon_N}.  

Eqs.~(6) and (8) of Ref.~\cite{FPhys} show that the operator on the right side of \eq{eq:Upsilon_N in terms of H_F H_B} can be expanded and reordered using the Zassenhaus formula \cite{Suzuki} as follows
\begin{eqnarray}  \label{eq:Zassenhaus}
   &&\hspace{-5mm}\left[\exp(i\hat H_{\rm B}\delta t)+\exp(-i\hat H_{\rm F}\delta t)\right]^N\\
   &&\hspace{2mm}= \sum_{n=0}^N \exp[i\hat H_{\rm B}(N-n)\delta t]\exp(-i\hat H_{\rm F}n\delta t)\nonumber\\
   &&\hspace{2mm} \times\sum\limits_{v=0}^{N-n}
   \cdots \sum\limits_{\ell =0}^s \sum\limits_{k=0}^\ell \exp \left[
    (v+\cdots +\ell +k)(\delta t ^{2}[\hat H_{F} ,\hat H_{B}]+\hat Q) \right]\nonumber
\end{eqnarray}
where $\hat Q$ contains terms representing higher order commutators of the form $[[\hat H_{\rm B},\hat H_{\rm F}],\cdots]$.  It follows from \eq{eq:commutator}  that $\hat Q=0$ here.
Substituting \eq{eq:Zassenhaus} into  \eq{eq:Upsilon_N in terms of H_F H_B} and then simplifying the resulting expression using Eqs.~(B.14) and (B.15) in Appendix B of Ref.~\cite{FPhys} yields 
\begin{equation}
         \label{eq:Upsilon with interference fn}
        \ket{\Upsilon_\lambda}_N \propto \sum_{n=0}^N  I_{N-n,n}(\delta t^2\lambda)\exp[i\hat H_{\rm B}(N\!-\!n)\delta t]\exp[-i\hat H_{\rm F}n\delta t]\ket{\phi}
\end{equation}
where
\begin{equation}
    \label{eq:interference function}
    I_{N-n,n}({z})
          =\exp[-in(N-n){z}/2]\prod_{q=1}^n \frac{\sin [(N+1-q){z}/2]}{\sin (q{z}/2)}
\end{equation}
is an \textit{interference function} that takes account of the non-commutativity of $\hat H_{\rm F}$ and $\hat H_{\rm B}$.

To relate this to what an observer in the galaxy would see, imagine that the galaxy contains a clock that is constructed from T-invariant matter.  We will refer to any time shown by the clock as ``clock time'' and use the symbol $t_{\rm c}$ to represent its value.  Let the state $\ket{\phi}$ represents the clock showing the time $t_{\rm c}=0$. The state
\begin{equation}  \label{eq:clock state}
     \exp[i\hat H_{\rm B}(N-n)\delta t]\exp[-i\hat H_{\rm F}n\delta t]\ket{\phi}
\end{equation}
represents evolution by $\exp[-i\hat H_{\rm F}n\delta t]$ in one direction of time followed by $\exp[i\hat H_{\rm B}(N-n)\delta t]$ in the opposite direction which, by convention, first increases $t_{\rm c}$ by $n\delta t$ and then decreases it by $(N-n)\delta t$, respectively.  The state in \eq{eq:clock state} would therefore represent the clock showing the net clock time of
\begin{equation}  \label{eq:clock time}
t_{\rm c}=(2n-N)\delta t\ ,
\end{equation}
and so the state in \eq{eq:Upsilon with interference fn} represents a weighted superposition of states over the range of net clock times from $t_{\rm c}=-N\delta t$ to $N\delta t$.

\subsection{Time reversal invariance}

It is useful to first consider the special case where the universe is invariant under time reversal.  For this we set $\lambda=0$, $\hat H_{\rm F}=\hat H_{\rm B}=\hat H$ in \eq{eq:Upsilon with interference fn}. The interference function for $\lambda=0$ is the binomial coefficient $I_{N-n,n}(0)=\left({}^N_n\right)$ and so
\begin{equation}   \label{eq:temporal binomial state}
       \ket{\Upsilon_0}_N \propto  \sum_{n=0}^N B_n\exp[-i(2n-N)\hat H\delta t]\ket{\phi}
\end{equation}
where $B_n$ is given by \eq{eq:binomial coefficients}.  The coefficient $B_n$ becomes proportional to the Gaussian function $\exp[-(2n-N)^2/2N]$ for large $N$ and so
\[
       \ket{\Upsilon_0}_N \srel{\propto}{\sim}  \sum_{n=0}^N \exp[-(2n-N)^2/2N]\exp[-i(2n-N)\hat H\delta t]\ket{\phi}\ .
\]
Re-expressing the summation in terms of the index $m=2n-N$ and using the definition $\delta t=\sigma_{\rm t}/\sqrt{N}$ then yields
\begin{equation}   \label{eq:Upsilon_N temporal gaussian}
    \ket{\Upsilon_0}_N \srel{\propto}{\sim}  \sum_{m\in S} \exp[-\frac{(m\delta t)^2}{2\sigma^2_{\rm t}}]\exp(-i\hat H m\delta t)\ket{\phi}
\end{equation}
where $S=\{-N,-N+2,\ldots,N\}$. We define the large-$N$ limit as
\begin{equation}   \label{eq:temporal gaussian state}
    \ket{\Upsilon_0} = \lim\limits_{N\to\infty} \ket{\Upsilon_0}_N \propto  \int dt\ g(t,\sigma_{\rm t})\exp(-i\hat H t)\ket{\phi}
\end{equation}
where $g(t,\sigma_{\rm t})$ is given by \eq{eq:gaussian function}.  Although Fig. \ref{fig:TInv} is explicitly for the spatial case, it can also be used here as a comparison of $B_n$ and $g(t,\sigma_{\rm t})$ in Eqs.~(\ref{eq:temporal binomial state}) and (\ref{eq:temporal gaussian state}) provided we interpret the horizontal axis as $t/\sigma_{\rm t}$. Likewise, for  $N\ge N^{\rm (time)}_{\rm min}$ the locus of points representng $B_n$ is essentially equivalent to the curve $g(t,\sigma_{\rm t})$ up to a proportionality constant, and so every $\ket{\Upsilon_0}_N\in  \boldsymbol{\Upupsilon}_0$ in \eq{eq:set Upsilon} is physically indistinguishable from the state $\ket{\Upsilon_0}$ in \eq{eq:temporal gaussian state}.

Hence, for time reversal invariance, the construction yields a state, given by \eq{eq:temporal gaussian state}, that  
is a Gaussian weighted superposition of the time-translated states $\exp(-i\hat H t)\ket{\phi}$. This state represents the galaxy existing in time only for a duration of the order of $\sigma_{\rm t}$ and is analogous to \eq{eq:spatial gaussian state} which represents the centre of mass of the galaxy existing only in a spatial region with a size of the order of $\sigma_{\rm x}$.
Our construction, therefore, allows for the same kind of quantum state in time as in space, in the absence of T violation.
In other words, there is a symmetry between time and space for quantum states in this special case.
As discussed in the Introduction, this symmetry comes at the cost of the non-conservation of mass.

\subsection{Violation of time reversal invariance}

Next we examine the quite different situation of T violation where $\lambda\ne 0$ and $\hat H_{\rm F}\ne\hat H_{\rm B}$.
In that case the amplitudes for different virtual paths to the same point in time, as illustrated in Fig.~\ref{fig:random path}(b), can interfere leading to undulations in $I_{N-n,n}({z})$ as a function of $n$.  To find the values of $n$ where the modulus of the interference function $I_{N-n,n}(z)$ is maximized it is sufficient to look for the position where $|I_{N-n,n}(z)|$ is unchanged for consecutive values of $n$, i.e. where  $|I_{N-(n-1),n-1}(z)|=|I_{N-n,n}(z)|$. This condition reduces, on using \eq{eq:interference function} and performing some algebraic manipulation, to $|\sin[(N+1-n)z/2]|=|\sin(nz/2)|$.  Note that Eqs.~(\ref{eq:Upsilon with interference fn}) and (\ref{eq:interference function}) imply $z=\delta t^2\lambda$ and given $\delta t=\sigma_{\rm t}/\sqrt{N}$ from \eq{eq:time step}, this means $z$ is inversely proportional to $N$; thus we let $z=\theta/N$ where
\[
    \theta=\sigma_{\rm t}^2\lambda
\]
is the coefficient of proportionality (i.e. $\theta$ is independent of $N$).  Hence we wish to know the values of $n$ that satisfy $|\sin[(N+1-n)\theta/2N]|=|\sin(n\theta/2N)|$.  Writing $x=\theta(N+1)/2N$ and $y=n\theta/2N$ transforms this equation into $|\sin(x-y)|=|\sin(y)|$ which has the solutions $y=(x-\pi)/2+m\pi$ for integer $m$. Re-expressing the solutions in terms of $n$ then gives
\[
n=\frac{N+1}{2}+\frac{N(2m-1)\pi}{\theta}\ .
\]
The modulus of the interference function reaches a maximum value at this value of $n$ and one less (i.e for $n-1$).  Taking the midpoint and choosing the particular values $m=0, 1$ then gives the positions of two maxima (or ``peaks'') at $n=n_\pm$ where \begin{equation}   \label{eq:n for peaks}
   n_{\pm}=N\left( \sfc{1}{2} \pm\sfc{\pi}{\theta}\right)\ .
\end{equation}
Substituting $n_\pm$ for $n$ in \eq{eq:clock time} gives the corresponding clock times as
\begin{equation}  \label{eq:peak clock time}
    \pm t_{\rm c}^{\rm (peak)}=(2n_\pm-N)\delta t=\pm\frac{2\pi\sigma_{\rm t}\sqrt{N}}{\theta}
\end{equation}
where $t_{\rm c}^{\rm (peak)}$ is defined to be positive.

The modulus of the interference function \eq{eq:interference function} is shown in Section A of the Supplementary Material to be approximately Gaussian about these maxima, which allows us to write $\ket{\Upsilon_\lambda}_N$ in \eq{eq:Upsilon with interference fn} as a superposition of two states as follows:
\begin{equation} \label{eq:Upsilon with two peaks}
     \ket{\Upsilon_\lambda}_N \propto \ket{\Upsilon_\lambda^{(+)}}_N+\ket{\Upsilon_\lambda^{(-)}}_N
\end{equation}
where
\begin{equation}    \label{eq:Upsilon_N in terms of f and g}
        \ket{\Upsilon_{\lambda}^{(\pm)}}_N \propto \sum_{n=0}^N  f_{n}^{(\pm)} g_{n}^{(\pm)}\exp[i\hat H_{\rm B}(N\!-\!n)\delta t]\exp[-i\hat H_{\rm F}n\delta t]
    \ket{\phi}
\end{equation}
for $2\pi<\theta<4\pi$.  Here
\begin{eqnarray}
    \label{eq:f_n}
    f_{n}^{(\pm)}&=&\exp\{-i[n_+n_- - (n-n_\pm)^2]\theta/2N\}\ ,\\
    \label{eq:g_n}
    g_{n}^{(\pm)}&=&\exp[-(n-n_{\pm})^2|\theta\tan(\theta/4)|/2N]
\end{eqnarray}
are a complex phase function and a Gaussian weighting function, respectively.   Keeping in mind the definition of the clock time $t_{\rm c}$ from \eq{eq:clock time} for the state in \eq{eq:clock state}, we find that $\ket{\Upsilon_\lambda^{(\pm)}}_N$ is a Gaussian-weighted superposition of states over a range of clock times with a mean of $t_{\rm c}=\pm t_{\rm c}^{\rm (peak)}$ and a variance of $(\Delta t_{\rm c})^2 \approx 2/|\lambda\tan(\theta/4)|$.  In other words, the states $\ket{\Upsilon_\lambda^{(+)}}_N$ and $\ket{\Upsilon_\lambda^{(-)}}_N$ represent the universe localised in time for a duration of the order of $\Delta t_{\rm c}$ about the mean times $t_{\rm c}= t_{\rm c}^{\rm (peak)}$ and $t_{\rm c}= -t_{\rm c}^{\rm (peak)}$, respectively.

\begin{figure}  
 	\begin{center}
  		\includegraphics[width=9.0cm]{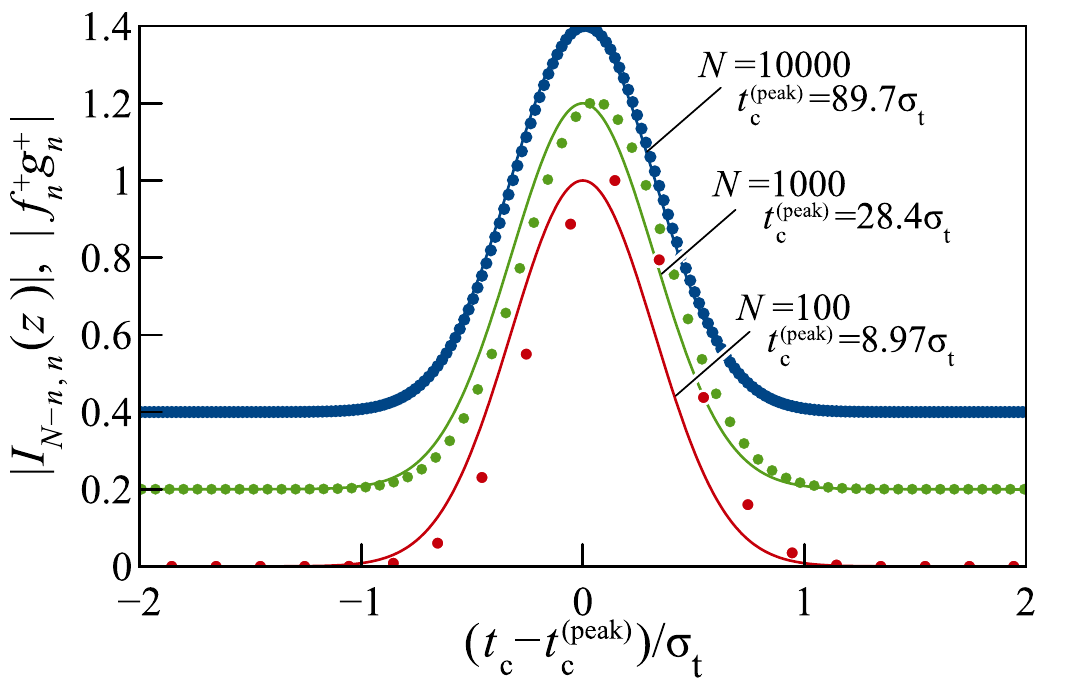}
 	\end{center}
 	\vspace{-8mm}
 	\caption{$|I_{N-n,n}(z)|$ plotted as a function of the scaled clock time $(t_{\rm c}-t^{\rm (peak)}_{\rm c})/\sigma_{\rm t}$ where $t_{\rm c}=(2n-N)\delta t$. The points $(|I_{N-n,n}(z)|,(t_{\rm c}-t^{\rm (peak)}_{\rm c})/\sigma_{\rm t})$ are generated parametrically by varying $n$.  The dots represent the exact values from \eq{eq:interference function} and the solid curves represent the approximation given by $|f_{n}^{+}g_{n}^{+}|$ in \eq{eq:Upsilon_N in terms of f and g}.  The numerical values used are $z=\theta/N$ where $\theta=2.23\pi$ and $N=100$ (red curve), $N=1000$ (green) and $N=10000$ (blue).  For clarity, the functions have been scaled to give a maximum of unity, and the green ($N=1000$) and blue ($N=10000$) curves have been displaced vertically by 0.2 and 0.4, respectively.
 		\label{fig:approx}}
\end{figure}    

The symmetry of the clock times associated with $\ket{\Upsilon_\lambda^{(+)}}_N$ and $\ket{\Upsilon_\lambda^{(-)}}_N$ about the time $t_{\rm c}=0$ reflects the symmetry of the construction \eq{eq:set Upsilon} and \eq{eq:Upsilon_N} which has no bias toward one direction of time or the other.  Moreover, if the state $\ket{\phi}$ is T invariant (i.e. if $\hat{\mathbf{T}}\ket{\phi}\propto\ket{\phi}$) and we shall assume that it is, then $\hat{\mathbf{T}}\ket{\Upsilon_\lambda^{(+)}}_N\propto\ket{\Upsilon_\lambda^{(-)}}_N$ and  $\hat{\mathbf{T}}\ket{\Upsilon_\lambda}_N\propto\ket{\Upsilon_\lambda}_N$.  This symmetry also arises in time-symmetric cosmological and gravitational studies of the direction of time \cite{Carroll,Barbour}. As the time evolution in one component of the superposition in \eq{eq:Upsilon with two peaks} is mirrored in the other, it suffices for us to consider just $\ket{\Upsilon_\lambda^{(+)}}_N$ and its corresponding value of $t_{\rm c}^{\rm (peak)}=2\pi\sqrt{N}\sigma_{\rm t}/\theta$.  Accordingly, we will call this value of $t_{\rm c}^{\rm (peak)}$  the \textit{representative clock time} and use it to label the whole state $\ket{\Upsilon_\lambda}_N$. The minimum representative clock time of a state in the set $\boldsymbol{\Upupsilon}_\lambda$ is found, using \eq{eq:peak clock time} with $N=N^{\rm (time)}_{\rm min}=\sigma_{\rm t}^2/\delta t_{\rm min}^2$ and $\theta = \sigma^2_{\rm t}\lambda$, to be
\begin{equation}  \label{eq:shortest time}
   t_{\rm c, min}^{\rm (peak)}=\frac{2\pi}{\lambda \delta t_{\rm min}}\ .
\end{equation}
A discussion of the values of $\lambda$ and $\delta t_{\rm min}$ in relation to $t_{\rm c, min}^{\rm (peak)}$ is given in Section C of the Supplementary Material.

Figure \ref{fig:approx} compares the coefficients $I_{N-n,n}(z)$ of the state $\ket{\Upsilon_\lambda}_N$ in \eq{eq:Upsilon with interference fn} with their Gaussian approximation $f_{n}^{+}g_{n}^{+}$ in \eq{eq:Upsilon_N in terms of f and g} near a maximum. The coefficients have been plotted as a function of $(t_{\rm c}-t_{\rm c}^{\rm (peak)})/\sigma_{\rm t}$ to centre them in the figure, where $t_{\rm c}^{\rm (peak)}$ is the position of the maxima given by \eq{eq:peak clock time}. As in Fig.~\ref{fig:TInv}, the values of $N$ have been chosen purposely to exaggerate the discreteness of the state $\ket{\Upsilon_\lambda}_N$. However, for every $\ket{\Upsilon_\lambda}_N\in \boldsymbol{\Upupsilon}_\lambda$ in \eq{eq:set Upsilon}, the values of $N$ are sufficiently large (i.e. $N\ge N^{\rm (time)}_{\rm min}$) that the locus of points representing  $I_{N-n,n}(z)$ is essentially equivalent to the Gaussian approximation  $f_{n}^{+}g_{n}^{+}$ up to a proportionality constant. It follows that each $\ket{\Upsilon_\lambda}_N\in \boldsymbol{\Upupsilon}_\lambda$ is physically indistinguishable from a state given by \eq{eq:Upsilon with two peaks} and \eq{eq:Upsilon_N in terms of f and g} with the same value of $N$ but where the sum over $n$ in \eq{eq:Upsilon_N in terms of f and g} is replaced with its integral equivalent.

\begin{figure}  
	\begin{center}
		\includegraphics[width=9.0cm]{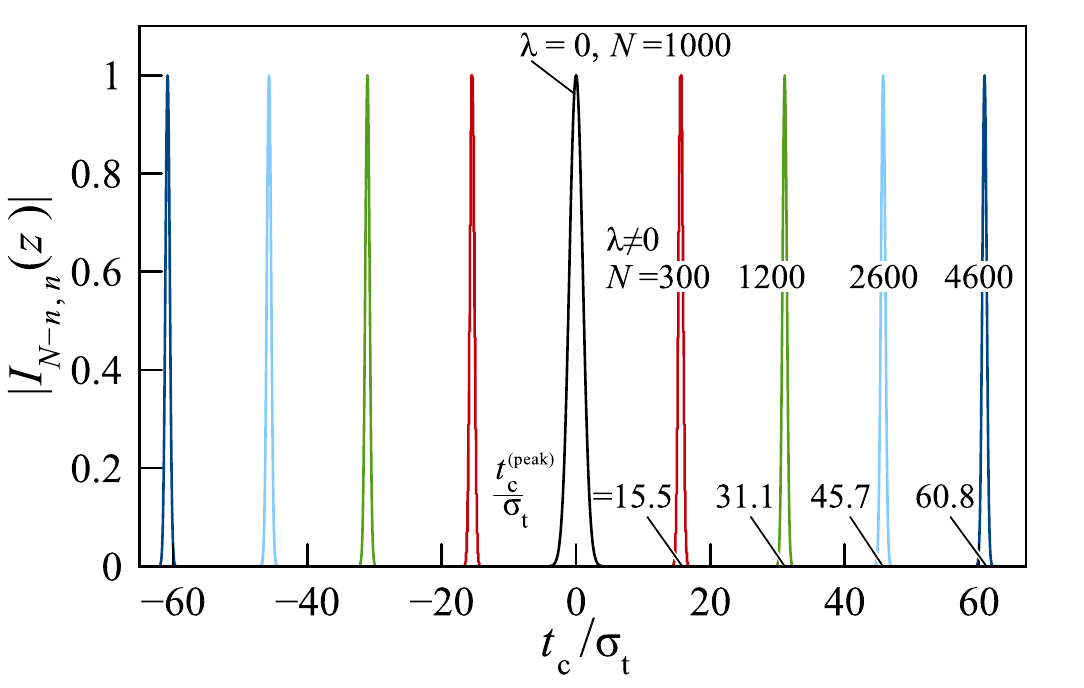}
	\end{center}
	\vspace{-5mm}
	\caption{$|I_{N-n,n}(z)|$ as a function of the scaled clock time $t_{\rm c}/\sigma_{\rm t}$ where $t_{\rm c}=(2n-N)\delta t$ for different values of $\lambda$ and $N$. As in Fig.~\ref{fig:approx}, the points $(|I_{N-n,n}(z)|,t_{\rm c}/\sigma_{\rm t})$ are generated parametrically by varying $n$.  For clarity, in each case straight lines connect consecutive discrete points of $|I_{N-n,n}(z)|$ to form a continuous curve.  The black curve represents the T invariant case (i.e. $\lambda=0$) and has been generated for $N=1000$.  It does not visibly change with increasing values of $N$.  The remaining curves represent the T violation case (i.e. $\lambda\ne 0$) for $\theta=2.23\pi$ and a range of $N$ values as follows: red curve for $N=300$ and $t_{\rm c}^{\rm (peak)}=15.5\sigma_{\rm t}$, green curve for $N=1200$  and $t_{\rm c}^{\rm (peak)}=31.1\sigma_{\rm t}$, light blue curve for $N=2600$ and $t_{\rm c}^{\rm (peak)}=45.7\sigma_{\rm t}$, and dark blue curve for $N=4600$  and $t_{\rm c}^{\rm (peak)}=60.8\sigma_{\rm t}$.  All curves have been scaled to give a maximum of unity.
		\label{fig:interference function}}
\end{figure}

The broad properties of the states  $\ket{\Upsilon_\lambda}_N$ are illustrated in Fig.~\ref{fig:interference function} which shows $|I_{N-n,n}(z)|$ plotted as a function of the scaled clock time $t_{\rm c}/\sigma_{\rm t}$.  The black curve corresponds to the time reversal invariance case where $\lambda=0$ (and so $\theta= \sigma^2_{\rm t}\lambda=0$).
All other curves correspond to the violation of time reversal invariance (i.e. $\lambda\ne 0$) and have been generated for $\theta=2.23\pi$ which gives the minimum uncertainty in energy and time (see Section B of the Supplementary Material for details).  The figure illustrates how the location of the maxima at $t_{\rm c}=\pm t_{\rm c}^{\rm (peak)}$ increases with $N$ as given by \eq{eq:peak clock time}.

For clarity, $|I_{N-n,n}(z)|$ is plotted in Fig.~\ref{fig:interference function} only for a select few values of $N$ for which the peaks in the corresponding curves are widely separated.  To see how close the peaks can be, consider the difference $\delta t_{\rm c}^{\rm (peak)}$ in the representative clock times $t_{\rm c}^{\rm (peak)}$ of states $\ket{\Upsilon_\lambda}_N$ with consecutive values of $N$, which is found from \eq{eq:peak clock time} to be
\[
\delta t_{\rm c}^{\rm (peak)}=\frac{2\pi\sigma_{\rm t}\sqrt{N+1}}{\theta}-\frac{2\pi\sigma_{\rm t}\sqrt{N}}{\theta}
\approx  \frac{\sigma_{\rm t}\pi}{\theta\sqrt{N}}
\]
for large $N$. Noting that $N \ge N^{\rm (time)}_{\rm min}=\sigma_{\rm t}^2/\delta t_{\rm min}^2$ gives $\delta t_{\rm c}^{\rm (peak)} \le  (\pi/\theta)\delta t_{\rm min}$ and as $2\pi < \theta < 4\pi$ we find
\[
\delta t_{\rm c}^{\rm (peak)} <  \frac{1}{2}\delta t_{\rm min}\ .
\]
Hence, for any given time $t>t_{\rm c, min}^{\rm (peak)}$, there is a state in the set $\mathbf{\Upsilon}_\lambda$ given by \eq{eq:set Upsilon} whose representative clock time $t_{\rm c}^{\rm (peak)}$ is equal to $t$ to within the resolution limit $\delta t_{\rm min}$.

Fig.~\ref{fig:interference function} clearly shows that the inclusion of the violation of time reversal invariance {\it dramatically} changes the set $\boldsymbol{\Upupsilon}_{\lambda}$ in \eq{eq:set Upsilon} from one containing elements that are physically equivalent (represented by the black curve), to one containing states that are diverging in time (other curves).
This striking outcome warrants careful consideration.   Both sets $\boldsymbol{\Upupsilon}_{\lambda=0}$ and $\boldsymbol{\Upupsilon}_{\lambda\ne 0}$ have the same mathematical construction given by \eq{eq:set Upsilon};  the striking difference we found between them is due solely to the phenomenological Hamiltonian and whether it respects T symmetry ($\lambda=0$) or not ($\lambda\ne 0$).  All the states in $\boldsymbol{\Upupsilon}_{\lambda=0}$ are physically equivalent to a unique state, $\ket{\Upsilon_0}$, which represents the galaxy as existing for one particular finite period in time.  This constitutes phenomenology associated with T symmetry.  
In contrast, with T violation there are infinitely-many different states in the set $\boldsymbol{\Upupsilon}_{\lambda\ne 0}$.  There is no physically-based reason to suppose that any of them has special significance and so, by default, all states in $\boldsymbol{\Upupsilon}_{\lambda\ne 0}$ have equal status in representing the state of the galaxy in time. This \textit{pluralism} constitutes phenomenology associated with T violation. That different states can equally represent the galaxy is not contradictory because each state represents the galaxy at a different representative clock time. In fact, the same pluralism is assumed in conventional quantum physics, and is the root of the asymmetry between time and space discussed in the Introduction.

\subsection{Impact for quantum states in time and space}

These remarkable results manifest a fundamental difference between quantum states in time and space.  All the states in the set $\boldsymbol{\Uppsi}$, irrespective of whether the discrete symmetries are obeyed or not, represent the galaxy existing only in a region of order $\sigma_{\rm x}$ near $x=0$. 
Likewise, all the states in the set $\boldsymbol{\Upupsilon}_{\lambda=0}$ associated with T symmetry represent the galaxy existing only for a duration of order $\sigma_{\rm t}$ near $t_{\rm c}=0$. The fact that the states in the  set $\boldsymbol{\Upupsilon}_{\lambda=0}$ don't conserve mass is testament to mass conservation not being an explicit property of the construction defined by \eq{eq:set Upsilon} and \eq{eq:Upsilon_N}.
But for a set $\boldsymbol{\Upupsilon}_{\lambda'}$ associated with T violation with $\lambda'\ne 0$, for any given time $t\ge t_{\rm c, min}^{\rm (peak)}$ we have just seen that there is a state $\ket{\Upsilon_{\lambda'}}_N\in\boldsymbol{\Upupsilon}_{\lambda'}$, whose representative clock time $t_{\rm c}^{\rm (peak)}$ is equal to $t$ to within the resolution limit $\delta t_{\rm min}$. In other words, the set $\boldsymbol{\Upupsilon}_{\lambda'}$ contains a state that represents the galaxy's existence at each corresponding moment in time.  That being the case, it would not be unreasonable to regard the set as representing a \textit{history} of the universe. It follows that the set $\boldsymbol{\Upupsilon}_{\lambda'}$ represents the persistence of the mass of the galaxy over the same period of time, in so far as the Hamiltonians $\hat H_F$ and  $\hat H_B$ conserve mass.  This raises a subtle point regarding conservation laws; while they may be due to deep principles (such as Noether's theorem) they are not manifested in quantum mechanics unless the state persists over a period of time.  The crucial point being that in conventional quantum mechanics, the persistence of the state is essentially \textit{axiomatic} and ensured by adopting a compliant dynamical equation of motion whereas here it  arises \textit{phenomenologically} as a property of the set of states $\boldsymbol{\Upupsilon}_{\lambda'}$. Finally, on comparing the two sets $\boldsymbol{\Upupsilon}_{\lambda=0}$ and $\boldsymbol{\Upupsilon}_{\lambda'\ne0}$ one could even venture to say that T violation, in effect, \textit{causes} the contents of the universe to be translated or, indeed, to \textit{evolve}, over an unbounded period of time.

\section{\label{sec:emergence}Emergence of conventional quantum mechanics}

\subsection{\label{sec:conventional qm}Coarse graining over time}

The spread of the state $\ket{\Upsilon_\lambda}_N$ along the time axis, as illustrated by the plots of $|I_{N-n,n}(z)|$ in Fig.~\ref{fig:interference function}, represents a significant departure from conventional quantum mechanics for which states are interpreted as having no extension in time.  Nevertheless, the conventional formalism can be recovered in the following way.  Imagine that observations of the galaxy are made with a resolution in time that is much larger than the width of the Gaussian weighting function $g_{n}^{\pm}$ in \eq{eq:Upsilon_N in terms of f and g}.  Under such coarse graining, the summation in \eq{eq:Upsilon_N in terms of f and g} can be replaced by the term corresponding to the maximum in $g_{n}^{\pm}$ and so, for example,
\[
\ket{\Upsilon_\lambda^{(+)}}_N \srel{\propto}{\sim} \exp[i\hat H_{\rm B}(N-n_+)\delta t]\exp(-i\hat H_{\rm F}n_+\delta t) \ket{\phi}\ .
\]
We can re-express this state in terms of its representative clock time, $t_{\rm c}^{\rm (peak)}$, which we shall shorten to $t_{\rm c}$ for brevity, as
\begin{equation}  \label{eq:Upsilon_+ coarse grained}
\ket{\Upsilon_\lambda^{(+)}}_N \srel{\propto}{\sim} \exp(i\hat H_{\rm B}t_{\rm c}a_-)\exp(-i\hat H_{\rm F}t_{\rm c}a_+) \ket{\phi}
\end{equation}
where $a_\pm=n_\pm/(n_+-n_-)$ and we have used $t_{\rm c}=(2n_+-N)\delta t=(n_+-n_-)\delta t$ and $N-n_+=n_-$. At this level of coarse graining, the time step $\delta t$ is effectively zero and  $t_{\rm c}$ is effectively a continuous variable.
Making use of the Baker-Campbell-Hausdorff formula \cite{Suzuki} in \eq{eq:Upsilon_+ coarse grained} yields
\begin{equation}
\ket{\Upsilon_\lambda^{(+)}}_N \srel{\propto}{\sim} \exp(\sfc{1}{2}ia_+a_-t_{\rm c}^2\lambda)\exp[-i(\hat H_{\rm F}a_+ - \hat H_{\rm B}a_-)t_{\rm c}] \ket{\phi}\ .
\end{equation}
The complex phase factor can be accommodated by transforming to a new state, $\ket{\widetilde\Upsilon(t_{\rm c})}$, as follows
\begin{equation} \label{eq:Upsilon_+ coarse grained transformed}
    \ket{\widetilde\Upsilon(t_{\rm c})}= \exp(-\sfc{1}{2}ia_+a_-t_{\rm c}^2\lambda)\ket{\Upsilon_\lambda^{(+)}}_N\propto \exp[-i(\hat H_{\rm F}a_+ - \hat H_{\rm B}a_-)t_{\rm c}] \ket{\phi}\ .
\end{equation}
On taking the derivative with respect to $t_{\rm c}$ we recover Schr\"{o}dinger's equation,
\begin{equation}   \label{eq:schrodingers equation}
    \frac{d}{dt_{\rm c}}\ket{\widetilde\Upsilon(t_{\rm c})} \srel{\propto}{\sim}
    -i(\hat H_{\rm F}a_+ - \hat H_{\rm B}a_-)\ket{\widetilde\Upsilon(t_{\rm c})}\ .
\end{equation}
Here, the coarse-grained Hamiltonian  $(\hat H_{\rm F}a_+ - \hat H_{\rm B}a_-)$  is a linear combination of $\hat H_{\rm F}$ and $\hat H_{\rm B}$ owing to the fact that the quantum virtual path involves contributions from both. 

Notice that the differential equation \eq{eq:schrodingers equation} does not depend on the state $\ket{\phi}$.  The ambiguity associated with loosely specifying $\ket{\phi}$ as being sharply defined in time does not play a role here. It is true that different choices for $\ket{\phi}$ will lead to different states $\ket{\widetilde\Upsilon(t_{\rm c})}$, but that is no concern when the goal is to show that conventional quantum mechanics is recovered.  The fact that \eq{eq:schrodingers equation} results for all allowed choices of $\ket{\phi}$ is all that is needed for this.

It is useful at this point to divide the galaxy into two non-interacting subsystems, one whose Hamiltonian $\hat H^{\rm (i)}=\hat{\mathbf{T}}^{-1}\!\hat H^{\rm (i)}\hat{\mathbf{T}}$ is T-invariant and the remainder whose Hamiltonian $\hat H_{\rm F}^{\rm (v)}=\hat{\mathbf{T}}^{-1}\!\hat H_{\rm B}^{\rm (v)}\hat{\mathbf{T}}\ne\hat H_{\rm B}^{\rm (v)}$ is T-violating;  in that case we can write
\begin{equation}  \label{eq:Separated Hamiltonian}
    \hat H_{\rm F}=\hat H^{\rm (i)}\otimes\hat{\boldsymbol 1}^{\rm (v)}+\hat{\boldsymbol 1}^{\rm (i)}\otimes\hat H_{\rm F}^{\rm (v)}\ ,\
     \hat H_{\rm B}=\hat H^{\rm (i)}\otimes\hat{\boldsymbol 1}^{\rm (v)}+\hat{\boldsymbol 1}^{\rm (i)}\otimes\hat H_{\rm B}^{\rm (v)}
\end{equation}
where the superscripts ``i'' and ``v'' label operators associated with the state space of the T-invariant and T-violating Hamiltonians, respectively, and $\hat{\boldsymbol 1}^{(\rm \cdot)}$ is an appropriate identity operator.  Equation \eq{eq:schrodingers equation} can then be rewritten as
\begin{equation}  \label{eq:schrodinger eqn for I and V}
    \frac{d}{dt_{\rm c}}\ket{\widetilde\Upsilon(t_{\rm c})} \srel{\propto}{\sim}
     -i(\hat H^{\rm (i)}\otimes\hat{\boldsymbol 1}^{\rm (v)}+\hat{\boldsymbol 1}^{\rm (i)}\otimes\hat H_{\rm phen}^{\rm (v)})\ket{\widetilde\Upsilon(t_{\rm c})}
\end{equation}
where $\hat H_{\rm phen}^{\rm (v)}=\hat H_{\rm F}^{\rm (v)}a_+ - \hat H_{\rm B}^{\rm (v)}a_-$ is the phenomenological Hamiltonian for the  T-violating subsystem.

It is straightforward to show that the commutator of $\hat H_{\rm phen}^{\rm (v)}=\hat H_{\rm F}^{\rm (v)}a_+ - \hat H_{\rm B}^{\rm (v)}a_-$  with its time reversed version is
\[
[\hat H_{\rm phen}^{\rm (v)}, \hat{\mathbf{T}}^{-1}\!\hat H_{\rm phen}^{\rm (v)}\hat{\mathbf{T}}] = -i\frac{\theta}{2\pi}\lambda
\]
which is $\theta/2\pi$ times the commutator $[\hat H_{\rm F}^{\rm (v)}, \hat H_{\rm B}^{\rm (v)}]$.  Thus, in principle, the commutation relation could be used to distinguish the phenomenological Hamiltonians $\hat H_{\rm phen}^{\rm (v)}$  and $ \hat{\mathbf{T}}^{-1}\!\hat H_{\rm phen}^{\rm (v)}\hat{\mathbf{T}}$ from the more elementary versions $\hat H_{\rm F}^{\rm (v)}$ and  $\hat H_{\rm B}^{\rm (v)}$.

\subsection{Conventional formalism and potential experimental tests}

These results are important because they not only show how the conventional formalism of quantum mechanics is recovered, but they also show how the construction introduced here may be verified experimentally.  To see this consider the following three points.  First, \eq{eq:schrodinger eqn for I and V} shows that the T-invariant subsystem behaves in accord with the conventional Hamiltonian $\hat H^{\rm (i)}$ with respect to clock time $t_{\rm c}$.  This means that conventional quantum mechanics is recovered for this subsystem. Second, \eq{eq:schrodinger eqn for I and V} shows that, due to the coarse graining, the role of the clock time $t_{\rm c}$ has been reduced from being a physical variable that describes the location and uncertainty of the galaxy with respect to time as illustrated in Fig.~\ref{fig:interference function}, to being simply a parameter that labels a different state in the set $\boldsymbol{\Upupsilon}_\lambda$ according to the time $t_{\rm c}=t_{\rm c}^{\rm (peak)}$ of the maximum in $g_{n}^{+}$.  Indeed, its demoted role is the very reason we are able to recover Schr\"{o}dinger's equation.  Third, any experiments involving T-violating matter that are performed by observers in the galaxy would give results that are consistent with \eq{eq:schrodinger eqn for I and V} and so they would provide evidence of the phenomenological Hamiltonian $\hat H_{\rm phen}^{\rm (v)}$ in the same way that experiments in our universe give evidence of the Hamiltonian associated with meson decay.  Any demonstration that $\hat H_{\rm phen}^{\rm (v)}$ differs from the more elemental Hamiltonians $\hat H_{\rm F}^{\rm (v)}$ and $\hat H_{\rm B}^{\rm (v)}$ represents a ``smoking gun'' for the construction introduced here. Of course, this specific result can not be used in practice because it applies to the simple model of T violation chosen here for its clarity rather than accuracy, and also because the present knowledge of T violating Hamiltonians is based on empirical results and so it is limited to the phenomenological version of the Hamiltonians.  More realistic models of the universe and T violating mechanisms may provide experimentally testable predictions, such as novel deviations from exponential decay for T violating matter or local variations in clock time.  But these are beyond the scope of the present work whose aim is to show, in the clearest way possible, how T violation may underlie differences between time and space.

\section{\label{sec:conclusion}Discussion}

We began by drawing attention to the asymmetry between time and space in conventional quantum theory where states are presumed to undergo continuous translation over time whereas there is no corresponding presumption about the state necessarily undergoing translations over space.  We set out to explore an alternate possibility by introducing a new quantum formalism that gives both space and time analogous quantum descriptions.  In developing the formalism, we paid particular attention to subtle mathematical details that play no significant role in conventional quantum mechanics. These details involve explicitly taking into account the P and T symmetry operations, translations of states in space and time, and fundamental limits of precision.  We incorporated them in a mathematical construction where quantum states are represented as a superposition of random paths in space or time.  We found that with no P or T symmetry violations, quantum states had analogous representations in space and time:  just as matter can be represented as existing only in a finite region of space,  it can also be represented as existing only for a finite interval of time. Clearly the price we pay for this symmetry is absence of the conservation of mass.   However with the violation of T symmetry, dramatic differences between the representation of quantum states in space and time arise through the quantum interference between different paths. The state (and the matter it describes) is found to persist over an unbounded range of time values. This result gives a new appreciation of conservation laws: while they may be due to deep principles they are not manifested unless the state persists over a sufficient period of time.  The Schr\"{o}dinger equation of conventional quantum mechanics, where time is reduced to a classical parameter, also emerges as a result of coarse graining over time. As such, T violation is seen in the new formalism as being responsible for fundamental differences between space and time in conventional quantum mechanics.

The new formalism may also help resolve other perplexing issues associated with space and time.  For example, the arrows of time indicate a preferred direction from ``past'' to ``future'' \cite{Price}, but there is no analogous preferred direction of space. The new formalism appears to offer a basis for understanding why.  Indeed the set of states in time, $\boldsymbol{\Upupsilon}_\lambda$ for $\lambda\ne 0$ in \eq{eq:set Upsilon}, has a natural order over time in the following sense.  First recall that our interpretation of \eq{eq:Upsilon_N in terms of H_F H_B} is that $\exp(-i\hat H_{\rm F} t)$ and $\exp(i\hat H_{\rm B} t)$ are associated with physical evolution in different directions of time, whereas the inverses $\exp(i\hat H_{\rm F} t)$ and $\exp(-i\hat H_{\rm B} t)$ are associated with the mathematical operations of rewinding that physical evolution.  Within this context, the coarse-grained state $\ket{\widetilde\Upsilon(t_{\rm c})}$ in \eq{eq:Upsilon_+ coarse grained transformed} is interpreted as resulting from evolution by $t_{\rm c}a_+$ in the positive direction of time and $t_{\rm c}a_-$ in the reverse direction, giving a net evolution of $t_{\rm c}(a_+-a_-)=t_{\rm c}$ in time from the state $\ket{\phi}$.  Correspondingly, the state $\ket{\widetilde\Upsilon(t'_{\rm c})}$ with $t'_{\rm c}>t_{\rm c}$ represents a \textit{more-evolved state} than $\ket{\widetilde\Upsilon(t_{\rm c})}$.  In fact writing
\begin{equation} \label{eq:Upsilon evolving}
    \ket{\widetilde\Upsilon(t'_{\rm c})}\propto  \exp[-i(\hat H_{\rm F}a_+ - \hat H_{\rm B}a_-)\Delta t] \ket{\widetilde\Upsilon(t_{\rm c})}
\end{equation}
where $\Delta t=t'_{\rm c}-t_{\rm c}>0$ shows that $\ket{\widetilde\Upsilon(t'_{\rm c})}$ evolves from $\ket{\widetilde\Upsilon(t_{\rm c})}$. One might be tempted to argue that we could equally well regard $\ket{\widetilde\Upsilon(t_{\rm c})}$ as evolving from $\ket{\widetilde\Upsilon(t'_{\rm c})}$ because
\begin{equation}   \label{eq:Upsilon rewinding}
     \ket{\widetilde\Upsilon(t_{\rm c})}\propto  \exp[i(\hat H_{\rm F}a_+ - \hat H_{\rm B}a_-)\Delta t]\ket{\widetilde\Upsilon(t'_{\rm c})}\ ,
\end{equation}
but doing so would be inconsistent with our interpretation of \eq{eq:Upsilon_N in terms of H_F H_B}.  According to that interpretation, \eq{eq:Upsilon rewinding} represents the mathematical \textit{rewinding} of the physical evolution represented by \eq{eq:Upsilon evolving}.  Note that the state $\ket{\widetilde\Upsilon(t_{\rm c})}$ is a coarsed-grained version of the component $\ket{\Upsilon_\lambda^{(+)}}_N$ of $\ket{\Upsilon_\lambda}_N$ in \eq{eq:Upsilon with two peaks}; an analogous argument also applies to the coarse-grained version of the other component $\ket{\Upsilon_\lambda^{(-)}}_N$, and thus to the whole state $\ket{\Upsilon_\lambda}_N$.  Hence, the set of states $\boldsymbol{\Upupsilon}_\lambda$ for $\lambda\ne 0$ are ordered by the degree of time evolution from the state $\ket{\phi}$.  This gives two preferred directions of time away from the origin of the time axis and so represents a \textit{symmetric arrow of time}.  Time-symmetric arrows have also been explored by Carroll, Barbour and co-workers \cite{Carroll,Barbour}.  In stark contrast, there is no analogous ordering for $\boldsymbol{\Uppsi}$ in \eq{eq:set Psi}, the set of states distributed over space.  Indeed, all the states in $\boldsymbol{\Uppsi}$ are physically indistinguishable. Also the ordering of the set  $\boldsymbol{\Upupsilon}_\lambda$ vanishes at $\lambda= 0$ which corresponds to T symmetry.  It appears, therefore, that T violation is also responsible giving time a direction (in the sense of orientating time away from the occurrence of $\ket{\phi}$).

In addition to these conceptual results, the new formalism was also found to have potential experimentally-testable consequences.  Indeed, for a subsystem associated with T violation, the formalism predicts that the experimentally-determined Hamiltonian, $\hat H_{\rm phen}^{\rm (v)}$ in  \eq{eq:schrodinger eqn for I and V}, will be different to the Hamiltonians, $\hat H_{\rm F}^{\rm (v)}$ or $\hat H_{\rm B}^{\rm (v)}$ in \eq{eq:Separated Hamiltonian}, associated with conventional quantum mechanics.  Further work is needed to develop feasible experiments for testing predicted departures from conventional theory like this. An experimental verification of the new formalism would have profound impact on our understanding of time.

In conclusion, the importance of Feynman's sums over paths for describing quantum phenomena is well beyond doubt \cite{Feynman}.  A distinctive feature of the quantum virtual paths in the new formalism is that they explicitly take into account the violation of T symmetry. The new formalism has the advantage of giving \textit{time and space an equal footing at a fundamental level} while allowing familiar differences, such as matter being localised in space but undergoing unbounded evolution in time, to arise \textit{phenomenologically} due to the fact that T violation is a property of  translations in time and not space.
As such, the violation of the discrete symmetries are seen to play a defining role in the quantum nature of time and space.

 I thank D.T. Pegg, H.M. Wiseman, M.J. Hall and T. Croucher for helpful discussions.

\newpage

\renewcommand\thesubsection{\Alph{subsection}}   
\renewcommand\theequation{\Alph{subsection}.\arabic{equation}}   
\renewcommand\thepage{Supplementary material --- \arabic{page}}   
\setcounter{page}{1}

\makeatletter 
   \@addtoreset{equation}{section} 
   \@addtoreset{equation}{subsection}
\makeatother


{\Large \bfseries{ \scalebox{0.8}{S\scalebox{0.8}{UPPLEMENTARY MATERIAL FOR}}}}

\begin{center} 

{\large \bfseries{ Quantum asymmetry between time and space}}\\
\ \\

{Joan A. Vaccaro}\\
\textit{Centre for Quantum Dynamics, Griffith University, Nathan 4111 Australia}

\end{center}	


\subsection{\label{app:shape of maxima}Approximate shape of the maxima} 
An approximate form of the interference function $I_{N-n,n}({z})$ in \eq{eq:interference function} near its maxima can be found by retaining terms of order $1/\sqrt{N}$ or larger as follows.  Substituting $n=n_\pm+k$, where $k$ is an integer, into \eq{eq:interference function} and using $N-n_\pm=n_\mp$ gives
\begin{eqnarray}  \label{eq:approx I at a maxima}
   I_{N-n,n}({z})
      = |I_{N-n_\pm,n_\pm}({z})|\exp[-i(n_\pm+k)(n_\mp-k){z}/2]
       \prod_{r=1}^{k} \frac{\sin [(N+1-r-n_\pm){z}/2]}{\sin[(r+n_\pm)z/2]}\ .
\end{eqnarray}
Next, substituting for $n_\pm$ and $z$ using \eq{eq:n for peaks} and $z=\theta/N$, respectively, using trigonometric identities and performing some algebraic manipulations eventually shows that the iterated product in \eq{eq:approx I at a maxima} can be written as
\begin{eqnarray} \label{eq:approx iterated product}
   \prod_{r=1}^{k} \frac{\sin [(N+1-r-n_\pm){z}/2]}{\sin[(r+n_\pm)z/2]}
   =(-1)^k\prod_{r=1}^{k} \frac{\cos(A)\cos(B)+\sin(A)\sin(B)}{\cos(A)\cos(C)-\sin(A)\sin(C)}
\end{eqnarray}
with $A=\theta/4$, $B=(r-1)\theta/2N$ and  $C=r\theta/2N$.  
Note that $k$ represents the number of steps in time, each of duration $\delta t=\sigma_{\rm t}/\sqrt{N}$, from the position of the maxima.  We want the approximation to be valid for a finite range around the maxima in the limit $N\to\infty$, and so we need $k$, and thus $r$, to vary over the range from $1$ to order $\sqrt{N}$.  It follows that  we can use the approximations $\cos(B)\approx 1$, $\cos(C)\approx 1$, $\sin(B)\approx B$ and $\sin(C)\approx C$ to first order in $1/\sqrt{N}$ in \eq{eq:approx iterated product}, and so
\[
  \prod_{r=1}^{k} \frac{\sin [(N+1-r-n_\pm){z}/2]}{\sin[(r+n_\pm)z/2]}
     \approx(-1)^k\prod_{r=1}^{k} \frac{1+B\tan(A)}{1-C\tan(A)}\ .
\]
As $B\approx C\approx r\theta/2N\ll 1$ to the same order of approximation, we can further approximate this as
\[
  \prod_{r=1}^{k} \frac{\sin [(N+1-r-n_\pm){z}/2]}{\sin[(r+n_\pm)z/2]}
       \approx(-1)^k\prod_{r=1}^{k} \exp\left[\frac{(2r-1)\theta}{2N}\tan\left(\frac{\theta}{4}\right)\right]\ ,
\]
and then, on evaluating the product on the right side, we eventually find
\[
    \prod_{r=1}^{k} \frac{\sin [(N+1-r-n_\pm){z}/2]}{\sin[(r+n_\pm)z/2]}
     \approx(-1)^k\exp\left[\frac{k^2\theta}{2N}\tan\left(\frac{\theta}{4}\right)\right]\ .
\]
Substituting into \eq{eq:approx I at a maxima} then gives 
\begin{eqnarray*}
  I_{N-n,n}({z})
    =|I_{N-n_\pm,n_\pm}({z})|
    \exp\left[-i(n_-n_+-k^2)\frac{\theta}{2N}\right]
              \exp\left[\frac{k^2\theta}{2N}\tan\left(\frac{\theta}{4}\right)\right]\ . 
\end{eqnarray*}
The right-most factor is a Gaussian function of $k$ provided $\theta\tan(\theta/4)$ is negative.  To ensure that this is the case we set $2\pi<\theta<4\pi$. It follows from $N-n_+=n_-$ and the symmetry property $I_{N-m,m}({z})=I_{m,N-m}({z})$ that $I_{N-n_+,n_+}({z})=I_{N-n_-,n_-}({z})$.  Thus, noting $k=n-n_\pm$, we find
\[
   I_{N-n,n}({z})\propto f_{n}^{(+)}g_{n}^{(+)}+f_{n}^{(-)}g_{n}^{(-)}
\]
where $f_{n}^{(\pm)}$ and $g_{n}^{(\pm)}$ are defined by \eq{eq:f_n} and \eq{eq:g_n}, respectively.  Substituting this result into \eq{eq:Upsilon with interference fn} then leads to \eq{eq:Upsilon with two peaks}. 

\subsection{\label{app:min uncert}Minimum uncertainty in energy and time} 
We have defined $t_{\rm c}$ via \eq{eq:clock state} and \eq{eq:clock time} as the time measured by clock devices that are constructed from T-invariant matter---this avoids any difficulties that might arise in defining clocks that are constructed from T-violating matter in general.  However, for the particular case here where $[\hat H_{\rm B}, \hat H_{\rm F}]=i\lambda$ there are no such difficulties and a clock constructed from  both T-invariant and T-violating matter will consistently register the same clock time $t_{\rm c}$ irrespective of the path, and the value of $t_{\rm c}$ will be the same as for a clock that is entirely constructed from T-invariant matter.  To see this, consider the two paths represented by $\hat A\hat B\ket{\phi}$ and $\hat B\hat A\ket{\phi}$ where $\hat A=\exp(-i\hat H_{\rm F}n\delta t)$ and $\hat B=\exp[i\hat H_{\rm B}(N-n)\delta t]$.  It is straightforward to show using the Baker-Campbell-Hausdorff formula that \cite{Suzuki}
\begin{equation}   \label{eq:appendix AB ket phi}
    \hat A\hat B\ket{\phi} = \exp[-i\lambda n(N-n)\delta t^2]\hat B\hat A\ket{\phi}
\end{equation}
and so both paths result in the same state apart from a complex phase factor. If we regard the whole universe as being a device that registers clock time $t_{\rm c}$ and if $\ket{\phi}$ represents  $t_{\rm c}=0$ then \eq{eq:appendix AB ket phi} implies that both $\hat A\hat B\ket{\phi}$ and $\hat B\hat A\ket{\phi}$ represent it registering the clock time $t_{\rm c}=(2n-N)\delta t$.  
The clock time $t_{\rm c}$ is therefore  representative of the whole universe in this case.

Although we do not have an operator corresponding to the clock time $t_{\rm c}$, we can still estimate the uncertainty in $t_{\rm c}$ for the state $\ket{\Upsilon_\lambda^{(\pm)}}_N$ using the following heuristic argument.  The sum over $n$ in \eq{eq:Upsilon_N in terms of f and g} means that, in addition to any intrinsic uncertainty in the time represented by clocks due to the state  $\ket{\phi}$, there is an additional contribution due to the finite width of the Gaussian weighting function $g_{n}^{(\pm)}$.  In fact, taking into account the linear relationship between $t_{\rm c}$ and $n$ given by $t_{\rm c}=(2n-N)\delta t$, the variance in possible clock time values $t_{\rm c}$ will be at least $(2\delta t)^2$ times the variance in $n$.  Thus we can bound the uncertainty in clock time as $(\Delta t_{\rm c})^2 \gtrsim 4(\Delta n)^2\delta t^2$ where $(\Delta n)^2=\overline{n^2}-{\overline{n}}^2=N/[2|\theta\tan(\theta/4)|]$ is the variance in $n$ associated with the Gaussian probability distribution $|g_{n}^{(\pm)}|^2$, and so using $\delta t=\sigma_{\rm t}/\sqrt{N}$ and $\theta=\sigma_{\rm t}^2\lambda$ we find
\begin{equation} \label{eq:appendix min var in time - T violation}
    (\Delta t_{\rm c})^2 \gtrsim \frac{2}{|\lambda\tan(\theta/4)|}\ .
\end{equation}  

The variance in \eq{eq:appendix min var in time - T violation} depends on the value of $\theta$. Rather than use any value in the allowed range $2\pi<\theta<4\pi$, it would be useful to have one that has a particular physical meaning.  One such value corresponds to minimal uncertainties in $t_{\rm c}$, $H_{\rm F}$ and $H_{\rm B}$.  The first step in finding it is to  use the Robertson-Schr\"{o}dinger uncertainty relation \cite{Robertson} for the Hamiltonians $\hat H_{\rm F}$ and $\hat H_{\rm B}$:
\[
	(\Delta H_{\rm F})^2 (\Delta \hat H_{\rm B})^2 \ge 
	\sfc{1}{4}|\ip{\{\hat H_{\rm F},\hat H_{\rm B}\}}-2\ip{\hat H_{\rm F}}\ip{\hat H_{\rm B}}|^2
	+ \sfc{1}{4}|\ip{[\hat H_{\rm F},\hat H_{\rm B}]}|^2
\]
where $\{\hat A,\hat B\}=\hat A\hat B+\hat B\hat A$ and  $(\Delta A)^2=\ip{\hat A^2}-\ip{\hat A}^2$  is the variance in $\hat A$.  As $[\hat H_{\rm B},\hat H_{\rm F}]=i\lambda$, the minimum of the right side
occurs when the covariance is zero:
\begin{equation}   \label{eq:covariance zero}
   \ip{\{\hat H_{\rm F},\hat H_{\rm B}\}}-2\ip{\hat H_{\rm F}}\ip{\hat H_{\rm B}}=0\ .
\end{equation}
Thus the minimum uncertainty is given by
\begin{equation}    \label{eq:HUP for H_F H_B}
    (\Delta H_{\rm F})^2 (\Delta H_{\rm B})^2 = \frac{|\lambda|^2}{4}\ .
\end{equation}
With no bias towards one direction of time or the other, there is correspondingly no bias towards one version of the Hamiltonian or the other and so we take the minimum uncertainty condition for the energy as 
\begin{equation}   \label{eq:minimum equal uncert in H_F and H_B}
   (\Delta H_{\rm F})^2=(\Delta H_{\rm B})^2=\frac{|\lambda|}{2}\ . 
\end{equation}

Next we need to determine the relationship between $\Delta t_{\rm c}$, $\Delta H_{\rm F}$ and $\Delta H_{\rm B}$.
Unfortunately, there has not been any previous study of the time-energy uncertainty relation for the case of T violation where two versions of the Hamiltonian operate, and it would be beyond the scope of this work to analyse it in detail here.  Nevertheless, we can glean some insight into the problem as follows.  Consider the operator defined by
\[
    \hat H = \sfc{1}{2}(\hat H_{\rm F}+\hat H_{\rm B});
\] 
it is straightforward to show that $\hat {\mathbf{T}}^{-1}\! \hat H \hat {\mathbf{T}}=\hat H$ and so  $\hat H$ is T invariant.  We know from the discussion of \eq{eq:appendix AB ket phi} that the state $\ket{\psi}=\exp(-i\hat H_{\rm F}t_1)\exp[i\hat H_{\rm B}t_2]\ket{\phi}$ represents the time $t_{\rm c}=t_1-t_2$.  
Rearranging using the Baker-Campbell-Hausdorff formula \cite{Suzuki} shows that $\ket{\psi}\propto\exp[-i\hat H_{\rm F}(t_1+t_2)]\exp[i\hat H 2 t_2]\ket{\phi}$ and so it follows that the state $\exp[i\hat H 2 t_2]\ket{\phi}$ represents the time $t_{\rm c}=-2t_2$.  Thus $\hat H$ is clearly a generator of translations in time. A similar argument shows that the operator $\sfc{1}{2}(\hat H_{\rm F}-\hat H_{\rm B})$ does not generate translations in time.  This implies that there is a meaningful uncertainty relation for the clock time $t_{\rm c}$ and $\hat H$.  The uncertainty in $\hat H$ is related to that of $H_{\rm F}$ and $H_{\rm B}$ by
\[
   (\Delta H)^2 = \sfc{1}{4}\left[ (\Delta H_{\rm F})^2+ (\Delta H_{\rm B})^2+
   \ip{\{\hat H_{\rm F},\hat H_{\rm B}\}}-2\ip{\hat H_{\rm F}}\ip{\hat H_{\rm B}}\right]\ ,
\]
and if \eq{eq:minimum equal uncert in H_F and H_B} holds, then so does \eq{eq:covariance zero} and we find
\begin{equation}   \label{eq:variance in H=(H_F+H_B)/2}
     (\Delta H)^2=\frac{|\lambda|}{4}\ .
\end{equation}
It is easy to calculate the product of variances in $t_{\rm c}$ and $\hat H$ for a state like \eq{eq:Upsilon_N temporal gaussian}.  In the limit $N\to\infty$ the sum over $m$ in \eq{eq:Upsilon_N temporal gaussian} becomes an integral over $t=m\delta t$ and so
\begin{equation}   \label{eq:temporal gaussian}
    \lim_{N\to\infty}\ket{\Upsilon_0}_N \propto \int dt \exp(-\frac{t^2}{2\sigma^2_{\rm t}})\exp(-i\hat H t)\ket{\phi}  
\end{equation}
which is the temporal analogy of \eq{eq:gaussian spatial translation}.
Replacing $\sigma^2_{\rm t}$ with $2(\Delta t_{\rm c})^2$ and performing the integral in \eq{eq:temporal gaussian} in the eigenbasis of $\hat H$ yields
\[
      \lim_{N\to\infty}\ket{\Upsilon_0}_N \propto \exp[-\hat H^2(\Delta t_{\rm c})^2]\ket{\phi}\ .  
\]
The state $\ket{\phi}$ is assumed to have a large variance in energy;  in the limit that $\ket{\phi}$ is a uniform superposition of  the eigenstates of $\hat H$, the probability distribution for $\hat H$ for the state on the right side becomes a truncated Gaussian \cite{Barr} with a variance of $(\Delta H)^2\approx (1-2/\pi)/4(\Delta t_{\rm c})^2$.  Hence an approximate energy-time uncertainty relation for this particular class of states is
\[
     (\Delta H)^2 (\Delta t_{\rm c})^2\approx\frac{1}{4}(1-\frac{2}{\pi})\ .
\]
We will presume that this result also applies to the states $\ket{\Upsilon_\lambda^{(\pm)}}_N$ in \eq{eq:Upsilon_N in terms of f and g} without significant modification.  In that case using \eq{eq:variance in H=(H_F+H_B)/2} to replace $(\Delta H)^2$ gives
\begin{equation}  \label{eq:min var in time - energy-time reln}
        (\Delta t_{\rm c})^2\approx\frac{1}{|\lambda|}(1-\frac{2}{\pi})\ .
\end{equation}
This gives the least uncertainty in clock time for the case where the uncertainty in energy is minimized; to be clear, the uncertainty in $\Delta t_{\rm c}$  can be smaller than that given by \eq{eq:min var in time - energy-time reln}  provided the uncertainty in energy is higher than the minimum represented by the equality in \eq{eq:HUP for H_F H_B}. Comparing \eq{eq:appendix min var in time - T violation} and \eq{eq:min var in time - energy-time reln} and keeping in mind that $2\pi<\theta<4\pi$ shows that the minimum uncertainty in energy and time is given by $\tan(\theta/4)\approx -2/(1-2/\pi)$, i.e. for $\theta\approx 2.23\pi$. 

Note that the uncertainty $\Delta t_{\rm c}$ is for each of the components $\ket{\Upsilon_\lambda^{(\pm)}}_N$ and not for the whole state $\ket{\Upsilon_\lambda}_N$ in \eq{eq:Upsilon with two peaks}. This uncertainty is appropriate from the point of view of an observer within the galaxy for whom the states $\ket{\Upsilon_\lambda^{(\pm)}}_N$ equally describe the state of the universe up to the symmetry given by $\hat{\mathbf{T}}\ket{\Upsilon_\lambda^{(+)}}_N\propto\ket{\Upsilon_\lambda^{(-)}}_N$.

\subsection{\label{app:quantifying T violation}Quantifying the T violation} 
The minimum representative clock time $t_{\rm c, min}^{\rm (peak)}$ for the set $\mathbf{\Upsilon}_\lambda$ defined in \eq{eq:shortest time} and the uncertainty in the clock time $\Delta t_{\rm c}$ defined in \eq{eq:appendix min var in time - T violation} are important physical parameters.  To estimate their values we need to quantify the minimum physically resolvable time given by $\delta t_{\rm min}$ and the degree of T violation represented by the value of $\lambda$.  The Planck time, $t_{\rm P}=5.4\times 10^{-44}$~s, is widely used as the minimum resolvable time and so we will adopt it here and set $\delta t_{\rm min}=t_{\rm P}$.  

Quantifying $\lambda$ is a rather more difficult.  One possibility is to assume that it has of the same order of magnitude as that of meson decay in our universe.  The eigenvalue spectrum of the (Hermitian) commutator  $i[\hat H_{\rm F},\hat H_{\rm B}]$ for meson decay has been estimated to have a Gausian distribution with a mean of zero and a standard deviation of $\sqrt{f}\times 10^{57}~{\rm s}^{-2}$ where $f$ is the fraction of the estimated $10^{80}$ particles in the visible universe that contribute to kaon-like T violation \cite{FPhys}.  The standard deviation serves as a physically-meaningful value of $\lambda$ and so we set $\lambda=\sqrt{f}\times 10^{57}~{\rm s}^{-2}$. Using \eq{eq:shortest time} with these values of $\lambda$ and $\delta t_{\rm min}$  then gives the minimum representative clock time as
\[
t_{\rm c, min}^{\rm (peak)} \approx f^{-1/2}\times10^{-13}~{\rm s}\ . 
\]
Thus \eq{eq:schrodingers equation} and \eq{eq:schrodinger eqn for I and V} describe the coarse-grained time evolution of the model universe from this time onwards.  The corresponding value of the uncertainty in the clock time $\Delta t_{\rm c}$ is, from \eq{eq:min var in time - energy-time reln},  
\[
\Delta t_{\rm c}\approx f^{-1/4}\times 10^{-29}~{\rm s}\ .
\] 

Another way to quantify $\lambda$ is to treat it as if its value is chosen by nature in order that the minimum representative clock time  is equal to the minimum time resolution, i.e. to make $t_{\rm c, min}^{\rm (peak)}=\delta t_{\rm min}$.  In that case we find, using \eq{eq:shortest time}, that
\begin{equation}
\lambda = \frac{2\pi}{\delta t_{\rm min}^2}
\end{equation}
which becomes $\lambda\approx 10^{87}~{\rm s}^{-2}$ for $\delta t_{\rm min}=t_{\rm P}$.  Then using \eq{eq:min var in time - energy-time reln} we find the corresponding uncertainty in the clock time is 
\begin{equation}
\Delta t_{\rm c}  \approx \sfc{1}{4}\delta t_{\rm min}\ .
\end{equation}
This represents the most extreme situation where \eq{eq:schrodingers equation} and \eq{eq:schrodinger eqn for I and V} describe the coarse-grained time evolution for all times from $t_{\rm c}=0$ and the uncertainty in clock time is undetectable. 

Finally, we should add that any non-zero value of $\lambda$ will give rise to the qualitative behaviour described in the main text.  However, according to \eq{eq:shortest time}, as the value of $\lambda$ approaches zero, the minimum representative clock time $t_{\rm c, min}^{\rm (peak)}$ becomes correspondingly large and so the desired behaviour is confined to ever larger times $t_{\rm c}^{\rm (peak)}$.

\end{document}